\UseRawInputEncoding
\newif\ifreport
\reporttrue   

\documentclass[conference]{IEEEtran}
\IEEEoverridecommandlockouts

\AtBeginDocument{
  }

\usepackage{bm}
\usepackage{amsmath}
\usepackage{amssymb}
\usepackage{soul}
\usepackage{algorithm}
\usepackage{algpseudocode}
\usepackage[utf8]{inputenc}
\usepackage{graphicx}
\usepackage{cuted}
\usepackage[dvipsnames]{xcolor}
\usepackage{subcaption}

\newcommand{\ignore}[1]{}
\usepackage{color}

\def\blue{\color{blue}}

\newtheorem{theorem}{Theorem}[section]

\newtheorem{proposition}[theorem]{Proposition}

\newenvironment{proof}{\IEEEproof}{\endIEEEproof}

\begin{document}

\title{Significance-Driven Semantic Communication}

\author{
\IEEEauthorblockN{Christian McDowell$^*$, Andrea Panebianco$^*$, Sirin Chakraborty, and Yin Sun}
\IEEEauthorblockA{Dept. ECE, Auburn University\\
Auburn, Alabama, USA\\
}
\thanks{$^*$Both authors contributed equally to this research. 
This work was supported in part by the NSF under Grant CNS-2239677, the Alabama Research and Development Enhancement Fund (ARDEF) program under Grant Agreement No. 1ARDEF26 11,  a donation from NVIDIA, and Auburn University.
}
}

\maketitle

\begin{abstract}
In this paper, we study a significance-driven cross-layer semantic communication design problem. Based on statistical decision theory, we introduce an information-theoretic measure of per-sample data significance that quantifies the task-specific value of each individual observation. Using this metric, we formulate a cross-layer optimization problem that simultaneously optimizes (i) physical-layer semantic encoding and inference and (ii) MAC-layer resource allocation, with the objective of maximizing semantic spectrum efficiency, defined as the semantic value delivered per unit bandwidth per unit time. At the physical layer, we develop Meta-Learning Variational Information Bottleneck (Meta-VIB), a new semantic transceiver that employs a meta-learned hypernetwork to compress high-dimensional observations into semantically significant latents, enabling instantaneous adaptation to dynamic channel conditions and varying symbol budgets without online retraining. At the MAC layer, we model channel 
allocation as a Multi-Action Restless Multi-Armed Bandit (MA-RMAB) and adopt the $Q$-Maximization algorithm, which dynamically allocates channel resources to sensors based on their semantic value of information. Experimental results on a real-world pedestrian safety dataset demonstrate that our joint design achieves substantial gains in semantic spectrum efficiency over baselines, \textcolor{black}{reaching up to $1000$ times gain at an average SNR of 0 dB and $40$ times gain at an average SNR of 5 dB. 
}
\end{abstract}

\begin{IEEEkeywords}
Semantic Communication, Data Significance, Age of Information, Restless Bandit.
\end{IEEEkeywords}

\section{Introduction} \label{sec:intro}

The term ``semantics'' originates from the ancient Greek word $\sigma\eta\mu\alpha\nu\tau\iota\kappa\acute{o}\varsigma$ (semantikós), 
meaning ``significance.'' In edge intelligent systems, such as autonomous driving, industrial automation, public safety, and emergency response, the semantic importance of information is determined by its usefulness for downstream inference and decision-making.
In these high-stakes environments, sensors, robots, and automated systems must react rapidly to unexpected events while operating under stringent communication, computation, and energy constraints.
From this perspective, the goal of communication is not merely to deliver more data, but to deliver the \emph{right} data: the information that is most critical for downstream inference, control, and safety.


Despite its intuitive appeal, a rigorous theory for quantifying the value of individual data samples remains largely elusive. 
Without such a framework, many existing communication systems rely on indirect heuristics, such as data freshness or data type, rather than task-specific data value when making transmission decisions. This disconnect limits the network's ability to identify and
prioritize the information most relevant to downstream tasks. As a result, communication is often designed as a high-throughput data delivery pipeline, rather than as a significance-aware mechanism for extracting, compressing, and transmitting task-critical content.

This gap raises three fundamental questions: (i) How can we formally define and quantify the semantic value of sensed data with respect to a task-specific loss? (ii) How can a sensor dynamically extract and compress only the semantically significant information while adapting to fluctuating channel conditions and resource constraints?
 (iii) How should channel resources be allocated across competing sensors to maximize the overall semantic value delivered by the network? To address these challenges, we introduce a \emph{significance-driven semantic communication}  cross-layer design that jointly optimizes (i) physical-layer semantic encoding and inference and (ii) Medium Access Control (MAC)-layer resource allocation. Our design is guided by a \emph{per-sample data significance metric} and aims to maximize \emph{semantic spectrum efficiency}, defined as the task-relevant value delivered per unit bandwidth per unit time. The main technical contributions of this paper are summarized as follows:

\begin{itemize}


\item Based on statistical decision theory, we introduce an informa-tion-theoretic measure of per-sample data significance 
that quantifies the value of each individual observation under a given task loss function (Section~\ref{sec:inter}). Using this metric, we formulate a cross-layer semantic communication design problem that jointly optimizes (i) physical-layer encoding and inference and (ii) MAC-layer channel resource allocation to maximize semantic spectrum efficiency (Section~\ref{sec:SemanticCom}).


\item To solve this cross-layer design problem, at the physical layer we develop a \emph{Meta-learning Variational Information Bottleneck (Meta-VIB)} transceiver that generalizes across time-varying channel conditions and codelength constraints without requiring online retraining. Meta-VIB integrates Feature-wise Linear Modulation (FiLM) conditioning, codelength truncation, an information-concentrating regularizer, and a carefully designed training procedure to enable robust adaptation.
At the MAC layer, we employ the \emph{$Q$-Maximization} algorithm in \cite{chakraborty2026advantage} to dynamically allocate channel resources across
sensors based on the semantic value of each message (Section \ref{sec:cross-layer}).
To the best of our knowledge, this is \emph{the first cross-layer semantic 
communication design that jointly optimizes physical-layer semantic compression and MAC-layer resource allocation using a per-sample data significance metric.}

\item We validate our cross-layer design using a pedestrian safety dataset collected from live traffic-camera feeds in downtown Auburn, Alabama. Meta-VIB achieves up to a $10
00$ times improvement in average significance at a Signal-to-Noise Ratio (SNR) of $0$ dB, and up to a $10$ times improvement at a SNR of $5$ dB, compared with state-of-the-art semantic communication baselines. In addition, the (Meta-VIB, $Q$-Maximization) cross-layer design attains up to $1000$ times improvement in semantic spectrum efficiency at an  average SNR of $0$ dB and up to $40$ times improvement at an  average SNR of $5$ dB, compared with cross-layer baselines (Section \ref{sec:evaluation}).
\end{itemize}

\section{Related Work} 

At the physical layer, a variety of semantic communication designs have been proposed to enhance task-oriented performance.  Joint Source-Channel Coding (JSCC) and DeepJSCC  \cite{bourtsoulatze2019deep, xu2023deep, qiao2025todma} use deep learning to map source data directly to channel symbols. More recently, Variational Information Bottleneck (VIB)-based designs \cite{pinheiro2021variational, kountouris2021semantics, diao2025aligning, peng2025hyper} have emerged to learn compact, task-relevant representations. However, these designs typically utilize fixed latent dimensions and lack the flexibility to adapt to dynamic codelength constraints. Naively addressing such variability would require separate models or repeated retraining across different operating conditions, incurring high computational complexity. To overcome this, Meta-VIB employs meta-learning  to adapt to both time-varying SNR and dynamic codelengths without requiring online retraining.

At the MAC layer, Age of Information (AoI) has become a standard metric for quantifying information freshness and guiding resource scheduling \cite{sun2017update, yates2021age, fountoulakis2023scheduling, ramakanth2024monitoring, chakraborty2025send}. While effective in capturing timeliness, AoI does not account for semantic relevance. To address this limitation, several alternative metrics have been proposed, including Age of Incorrect Information (AoII) \cite{maatouk2020age}, Urgency of Information (UoI) \cite{zheng2020urgency}, AoI at Query (QAoI) \cite{holm2021freshness}, Uncertainty of Information \cite{chen2024optimal}, Value of Information (VoI) \cite{soleymani2024consistency}, among others. On the algorithmic side, traditional Whittle-index and LP-based policies~\cite{whittle1988restless, weber1990index, gast2024linear} typically assume binary actions, finite state spaces, or indexability. Conversely, existing Reinforcement Learning (RL) methods for restless multi-armed bandits~\cite{nakhleh2021neurwin, killian2021q, avrachenkov2020whittle, xiong2022index} 
often assume finite state spaces and require retraining when system configurations, such as the number of sensors or the available bandwidth, differ between training and deployment. Moreover, much of the above mentioned MAC-layer literature abstracts away physical-layer design considerations. To address these limitations, we adopt the low-complexity $Q$-Maximization algorithm~\cite{chakraborty2026advantage}, which schedules channel resources according to per-sample data significance in high-dimensional continuous state spaces and generalizes to various system configurations without retraining.

While the existing literature predominantly treats physical-layer semantic communication design and MAC-layer scheduling as isolated problems, our work unifies these layers. By leveraging an information-theoretic measure of per-sample significance, we formulate and solve a cross-layer semantic communication  problem.

\section{A Per-Sample Metric for Data Significance}
\label{sec:data_significance}

In this section, we revisit a \emph{per-sample} value-of-information metric from statistical decision theory and derive its information-theoretic interpretation. Specifically, we show that this metric can be expressed as a loss-based divergence, which provides a rigorous foundation for quantifying the significance of individual data samples.

\subsection{Preliminaries: Decision-Theoretic Foundation}

Consider the problem of estimating a target random variable $Y \in \mathcal{Y}$. A decision-maker selects an action $a$ from a feasible action space $\mathcal{A}$. The quality of that action is quantified by a loss function $L: \mathcal{Y} \times \mathcal{A} \to \mathbb{R}$, where $L(y,a)$ denotes the cost incurred when action $a$ is selected and the realized target value is $Y = y$. 

The choice of loss function $L$ is task-dependent and should be designed according to the estimation objective. For example, in minimum mean-squared error (MMSE) estimation, the action is a point estimate $\hat{y}$, and the loss is $L_2(y,\hat{y})= \|y-\hat{y}\|^2_2$. In maximum likelihood (ML) estimation, the action is a candidate distribution $Q_Y$ for $Y$, and the loss is the negative log-likelihood $L_{\log}(y, Q_Y) = -\log Q_Y(y)$, which is known as the \emph{logarithmic loss} \cite{grunwald2004game,farnia2016minimax,shisher2024timely}. In Section \ref{sec:evaluation}, we introduce a loss function tailored to a pedestrian safety monitoring application.

Suppose the decision-maker observes a realization $z$ of a random variable $Z$ that carries information about $Y$. This observation updates the prior distribution $P_Y$ to the posterior distribution $P_{Y|Z=z}$. Without observing $Z=z$, the decision-maker selects the prior-optimal action
\begin{equation}
    a^* = \arg\min_{a \in \mathcal{A}} \mathbb{E}_{Y \sim P_Y} [L(Y, a)],
\end{equation}
and the resulting minimum expected loss is called the \emph{Bayes risk} \cite{BlackwellGirshick1954} under the prior $P_Y$. 
After observing $Z=z$, the decision-maker instead chooses the posterior-optimal action
\begin{equation}
    a^*_z = \arg\min_{a \in \mathcal{A}} \mathbb{E}_{Y \sim P_{Y|Z=z}} [L(Y, a)],
    \label{eq_posterior_action}
\end{equation}
and the corresponding minimum expected loss is the Bayes risk under the posterior \(P_{Y|Z=z}\).
The reduction in expected loss induced by observing $Z=z$, for the purpose of predicting target $Y$, is called the \emph{conditional value of sampled information} \cite[Section 4.5.2]{raiffa1961applied}:
\begin{equation}
    v_{L,Y}(z) = \mathbb{E}_{Y \sim P_{Y|Z=z}} [L(Y, a^*)] - \mathbb{E}_{Y \sim P_{Y|Z=z}} [L(Y, a^*_z)],
    \label{eq:cvi_definition}
\end{equation}
which is the expected loss of the prior-optimal action $a^*$ under the posterior distribution $P_{Y|Z=z}$ minus that of the posterior-optimal action $a^*_z$ under the same posterior. Because $a^*_z$ is optimal under $P_{Y|Z=z}$, it follows immediately that $v_{L,Y}(z)\geq0$. 
Thus, $v_{L,Y}(z)$ quantifies how much the observation $Z=z$ improves decision-making for estimating the target $Y$. 

Averaging over all realizations of $Z$ yields the \emph{expected value of sampled information} \cite[Section 4.5.2]{raiffa1961applied}
\begin{align}
    \bar v_{L,Y,Z} &= \mathbb{E}_{Z\sim P_Z} [v_{L,Y}(Z)] \\
    &= \mathbb{E}_{Y \sim P_{Y}} [L(Y, a^*)] - \mathbb{E}_{(Y,Z) \sim P_{Y,Z}} [L(Y, a^*_Z)],
    \label{eq:ecvi_definition}
\end{align}
which measures the average benefit of observing \(Z\) for estimating the target $Y$.

\subsection{Information-Theoretic Interpretation}\label{sec:inter}
To make this framework operational for semantic communication, we now provide an information-theoretic interpretation of \(v_{L,Y}(z)\). 

To that end, we first introduce the notions of \(L\)-entropy, \(L\)-conditional entropy, \(L\)-cross entropy, \(L\)-divergence, and \(L\)-mutual information \cite{grunwald2004game,farnia2016minimax,shisher2024timely}. 
For a loss function \(L\), the \emph{\(L\)-entropy} of a random variable \(Y\) is defined as
\begin{equation}\label{eq_entropy}
    H_L(Y) = \min_{a \in \mathcal{A}} \mathbb{E}_{Y \sim P_Y}[L(Y,a)],
\end{equation}
which is the minimum expected loss incurred when predicting \(Y\) without any side information. In other words, $H_L(Y)$  is the Bayes risk under the prior distribution \(P_Y\).
In addition, for a realization \(Z=z\), the \emph{\(L\)-conditional entropy} of \(Y\) given \(Z=z\) is
\begin{equation}\label{eq:conditional_entropy}
    H_L(Y|Z=z) = \min_{a \in \mathcal{A}} \mathbb{E}_{Y \sim P_{Y|Z=z}}[L(Y,a)],
\end{equation}
and its average over $Z$ is
\begin{equation}\label{eq:avg_conditional_entropy}
    H_L(Y|Z) = \mathbb{E}_{Z \sim P_Z}[H_L(Y|Z=z)].
\end{equation}
Thus, \(H_L(Y|Z)\) is the average posterior Bayes risk.

Let \(a_{Q_Y}\) denote a \emph{Bayes action} \cite{BernardoSmith2000} minimizing the expected loss under distribution \(Q_Y\). The \emph{\(L\)-cross entropy} between two distributions \(P_Y\) and \(Q_Y\) is defined as
\begin{equation}\label{eq_cross}
    H_L(P_Y;Q_Y) = \mathbb{E}_{Y \sim P_Y}[L(Y,a_{Q_Y})].
\end{equation}
From \eqref{eq_entropy} and \eqref{eq_cross}, we can get $H_L(Y)=H_L(P_Y;P_Y) \leq H_L(P_Y;Q_Y)$ for all $Q_Y$.
The \emph{\(L\)-divergence} is then defined as
\begin{align}
    D_L(P_Y\|Q_Y) &
    = H_L(P_Y;Q_Y) - H_L(Y) \nonumber\\
    &= \mathbb{E}_{Y \sim P_Y}[L(Y,a_{Q_Y})] - \mathbb{E}_{Y \sim P_Y}[L(Y,a_{P_Y})],
\end{align}
which is non-negative.
Thus, \(D_L(P_Y\|Q_Y)\) measures the excess expected loss incurred when one uses the Bayes action optimized for \(Q_Y\) instead of the Bayes action optimized for the true distribution \(P_Y\). When \(L\) is the logarithmic loss, \(D_L(P_Y\|Q_Y)\) reduces to the classical Kullback-Leibler divergence \cite{grunwald2004game,farnia2016minimax,shisher2024timely}.

The \emph{\(L\)-mutual information} between \(Y\) and \(Z\) is defined as
\begin{equation}\label{eq:mutual_info1}
    I_L(Y;Z) = \mathbb{E}_{Z \sim P_Z}\!\left[D_L\!\left(P_{Y|Z=z}\,\|\,P_Y\right)\right].
\end{equation}
Equivalently,
\begin{equation}
    I_L(Y;Z) = H_L(Y) - H_L(Y|Z), \label{eq:mutual_info}
\end{equation}
which represents the average reduction in Bayes risk achieved by observing \(Z\). Under the logarithmic loss $L_{\log}$, $I_{L_{\log}}(Y;Z)$ reduces to the Shannon mutual information $I(Y;Z)$ and is therefore symmetric, i.e., $I_{L_{\log}}(Y;Z) = I_{L_{\log}}(Z;Y)$. In general, however, \(I_L(Y;Z) \neq I_L(Z;Y)\).

The following proposition establishes a direct bridge between decision theory and information theory by showing that the conditional value of sampled information is exactly an $L$-divergence.

\begin{proposition}[Information-theoretic interpretation of \(v_{L,Y}(z)\)]\label{Prop_info}
For any loss function $L$, target $Y$, and  realization \(z\) of \(Z\), 
\begin{equation}
    v_{L,Y}(z) = D_L\!\left(P_{Y|Z=z}\,\|\,P_Y\right).
\end{equation}
Moreover, taking the expectation over $Z$ yields
\begin{equation}
    \bar v_{L,Y,Z} = \mathbb{E}_{Z \sim P_Z}\!\left[D_L\!\left(P_{Y|Z=z}\,\|\,P_Y\right)\right] = I_L(Y;Z).
\end{equation}
\end{proposition}

\begin{proof}
The proof is straightforward; see
\ifreport
Appendix \ref{App_Prop_info}.
\else
our technical report \cite{mcdowell2026significance}.
\fi
\end{proof}

Proposition \ref{Prop_info} shows that \(v_{L,Y}(z)\) is precisely the  \(L\)-divergence between the posterior distribution $P_{Y|Z=z}$ and the prior distribution $P_Y$, while its expectation \(\bar v_{L,Y,Z}\) is the corresponding \(L\)-mutual information. This identity provides a rigorous interpretation of semantic significance: the value \(v_{L,Y}(z)\)  of an individual sample \(Z=z\) is exactly the excess expected loss incurred when that sample is discarded and decisions revert to the prior-optimal action for predicting $Y$. Because the semantic value \(v_{L,Y}(z)\) is defined through a task-specific loss function, its unit depends on the application and may represent, for example, avoided monetary loss, prevented physical damage, or lives saved as a result of observing  $Z=z$.

Therefore, \(v_{L,Y}(z)\) provides a task-aware, loss-based, information-theoretic measure of per-sample significance. In contrast to traditional transmission and scheduling formulations that focus only on the average value of information $\bar v_{L,Y,Z}$, e.g., \cite{shisher2024timely,shisher2023learning,KeyuanIEEEMASS2025,AriToN2026}, our work emphasizes the conditional quantity $v_{L,Y}(z)$, which captures the instantaneous value of a specific observation $Z=z$. This distinction is critical for our semantic communication framework in Sections \ref{sec:SemanticCom}-\ref{sec:evaluation}, where transmission and scheduling decisions are made at the level of individual samples rather than population averages.

\section{\!Significance-Driven Semantic Communication:
System Model and Problem Formulation}
\label{sec:SemanticCom}

This section presents the system model and  cross-layer
optimization problem for the proposed significance-driven semantic
communication framework. Our objective is to maximize the long-term
cumulative semantic significance delivered by the network.

\begin{figure*}[t]
 \centering
 \includegraphics[width=0.75\linewidth]{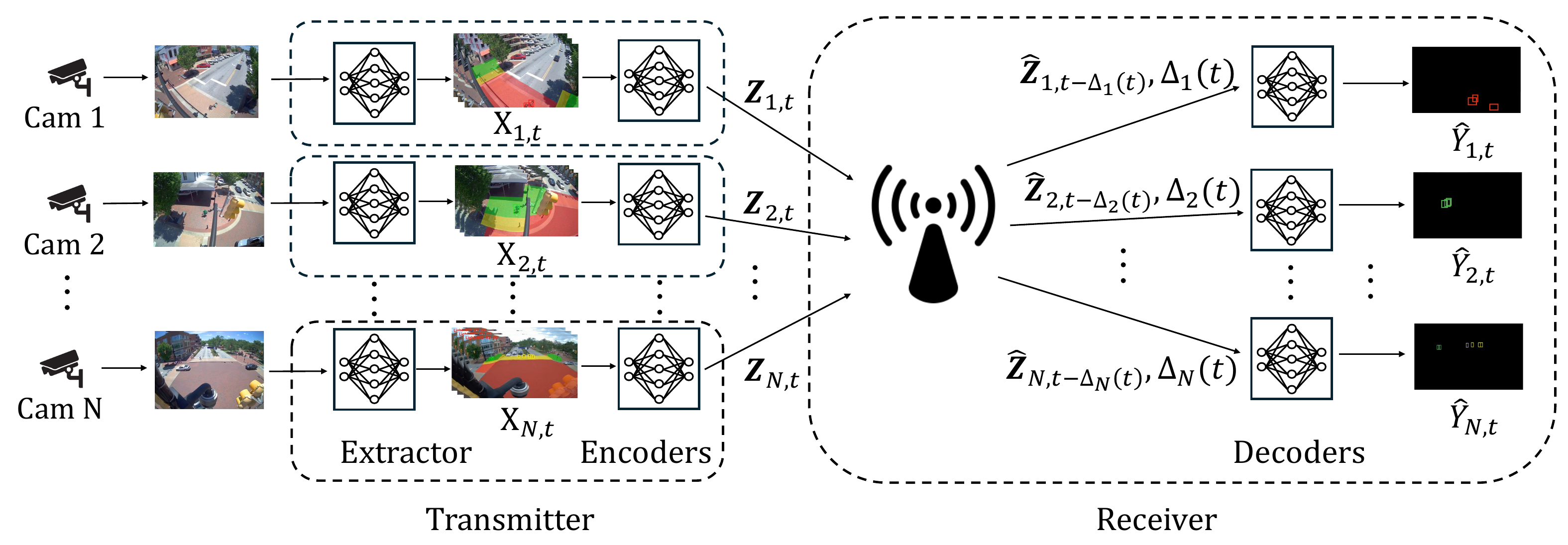}
 \vspace{-2mm}
 \caption{A significance-driven semantic communication system for pedestrian safety monitoring. 
 \ignore{
Each transmitter~$n$ runs YOLOv8s and DeepSORT
   to extract a semantic feature sequence~$X_{n,t}$, which the
   Meta-VIB encoder compresses into a latent
   codeword~$\mathbf{Z}_{n,t}$ for transmission over a
   multiple-access wireless channel. The receiver uses the
   time-stamped latent
   $S_{n,t}=(\hat{\mathbf{Z}}_{n,t-\Delta_n(t)},\,\Delta_n(t))$ to
   forecast pedestrian trajectories and safety labels. }}
 \label{fig:system_model}
 \vspace{-2mm}
\end{figure*}

\subsection{System Model}
\label{sec:system_model}

Consider a network of $N$ edge sensors transmitting to a common
receiver, as illustrated in Figure \ref{fig:system_model}. Time is slotted with slot duration $\tau$, and slots are
indexed by $t = 0, 1, 2, \ldots$. At slot $t$, sensor $n$ captures
a raw frame $F_{n,t}$ and applies a lightweight feature extractor
$f_n$ to obtain a semantic feature $V_{n,t} = f_n(F_{n,t})$.
To incorporate temporal context, sensor $n$ forms the historical feature
sequence $X_{n,t} = (V_{n,t}, V_{n,t-1}, \ldots, V_{n,t-u+1})$,
where $u$ is the history window length, chosen sufficiently large so
that $X_{n,t}$ captures all task-relevant temporal context needed for downstream
inference.

\emph{Encoder.}
Sensor~$n$ employs a semantic encoder~$e_n$, parameterized
by~$\phi_n$, that maps $X_{n,t}$ to a complex-valued latent codeword
$\mathbf{Z}_{n,t}=e_n(X_{n,t};\phi_n)\in\mathbb{C}^{\eta_{n,t}}$, where $\eta_{n,t}\in\{0,m,2m,\ldots,Km\}$ is the codelength (i.e., the number of complex channel symbols) allocated to sensor $n$ at slot $t$, with $m$ being
the basic symbol unit. If $\eta_{n,t}=0$, sensor $n$ remains silent.

\emph{Channel Model.}
The codeword~$\mathbf{Z}_{n,t}$ is transmitted over an orthogonal
multiple-access block-fading channel with  coefficient $h_{n,t}\in\mathbb{C}$. The received signal is
\begin{align}
  \hat{\mathbf{Z}}_{n,t}
    = h_{n,t} \mathbf{Z}_{n,t}+\boldsymbol{\varepsilon}_{n,t},
  \label{Gaussian_Channel}
\end{align}
where
$\boldsymbol{\varepsilon}_{n,t}\sim\mathcal{CN}(\mathbf{0},
\sigma^2\mathbf{I}_{\eta_{n,t}})$ is additive circularly symmetric complex Gaussian noise and $\mathbf{I}_n$ is the $n \times n$ identity matrix. Subject to the
power constraint
$\mathbb{E}[\|\mathbf{Z}_{n,t}\|_2^2\mid\eta_{n,t}]\le\eta_{n,t}P$,
the Shannon capacity per symbol is
\begin{align}\label{eq:capacity}
C(\zeta_{n,t})=\log_2(1+\zeta_{n,t}),
\end{align}
where 
$\zeta_{n,t}=|h_{n,t}|^2 P/\sigma^2$ is the SNR. Since
the wireless medium is shared, the aggregate number of channel
symbols allocated across all sensors at slot $t$ must satisfy
\begin{align}
  \sum_{n=1}^N \eta_{n,t}\le W,
\end{align}
where $W=\kappa B\tau$, $B$ is the channel bandwidth, and
$\kappa\in(0,1)$ is the
effective bandwidth fraction after accounting for protocol overhead.

\emph{Decoder.}
When sensor \(n\) is not scheduled, i.e., \(\eta_{n,t}=0\), the receiver retains the most recently received latent for inference. Let \(\hat{\mathbf{Z}}_{n,t-\Delta_n(t)}\) denote the freshest received latent from sensor \(n\), where \(\Delta_n(t)\) is its Age of Information (AoI) \cite{kaul2012real,yates2021age}, evolving as
\begin{align}\label{eq:aoi}
\Delta_n(t+1)=
\begin{cases}
1, & \text{if } \eta_{n,t}>0,\\
\Delta_n(t)+1, & \text{if } \eta_{n,t}=0.
\end{cases}
\end{align}
The freshest available latent \(\hat{\mathbf{Z}}_{n,t-\Delta_n(t)}\) together with its AoI \(\Delta_n(t)\) form the \emph{time-stamped latent} 
\begin{align}
S_{n,t} \triangleq (\hat{\mathbf{Z}}_{n,t-\Delta_n(t)}, \Delta_n(t)).
\end{align}
Let \(Y_{n,t}\) denote the task-specific target for sensor \(n\) inferred at the receiver in slot \(t\). 
A decoder \(g_n\), parameterized by \(\theta_n\), maps \(S_{n,t}\) to the estimate
 $a_{n,t} = g_n(S_{n,t};\theta_n)$, incurring loss $L_n(Y_{n,t},a_{n,t})$. The loss function $L_n$ is task-dependent and specifies the inference objective.

\emph{Scheduler.}
At each slot, the receiver determines the codelength allocation. An
admissible scheduling policy~$\pi$ maps receiver-side information to
$\boldsymbol{\eta}_t=(\eta_{1,t},\ldots,\eta_{N,t})$. The
information sets on which $e_n$, $g_n$, and $\pi$ operate are
specified next.

\subsection{Information Structure}
\label{sec:info_structure}

The encoder~$e_n$, predictor~$g_n$, and scheduler~$\pi$ operate on
distinct information sets, reflecting their different roles and
physical locations within the system.

\emph{Encoder.}
The encoder $e_n$ operates at sensor $n$ on the local state
$o_{n,t} \triangleq \left(X_{n,t},\, h_{n,t},\,
\eta_{n,t}\right)$,
where $X_{n,t}$ is the source sequence to compress, and $h_{n,t}$
and $\eta_{n,t}$ together determine the capacity budget available
for transmitting the latent codeword $\mathbf{Z}_{n,t}$.

\emph{Decoder.}
The decoder~$g_n$ operates at the receiver on
$(S_{n,t},h_{n,t}, $ $\eta_{n,t})$, where the channel
coefficient~$h_{n,t}$ enables coherent communication and $\eta_{n,t}$ specifies the
codelength of received noisy latent.

\emph{Scheduler.}
The scheduler~$\pi$ operates at the receiver on the aggregate  scheduling state
$\boldsymbol{\omega}_t\triangleq(\omega_{1,t},\ldots,\omega_{N,t})$, where $\omega_{n,t}\triangleq (S_{n,t},\zeta_{n,t})$. The scheduler requires $S_{n,t}$ because it is sample-based, and it requires
$\zeta_{n,t}$ because codelength allocation must
account for the per-symbol capacity~$C(\zeta_{n,t})$. Let $\Pi$
denote the class of all admissible
policies mapping ~$\boldsymbol{\omega}_t$ to the allocation
vector~$\boldsymbol{\eta}_t=(\eta_{1,t},\ldots,\eta_{N,t})$.

\subsection{Cross-Layer Optimization Problem and Semantic Spectrum
Efficiency}
\label{sec:joint_problem}

We jointly optimize (i) the physical-layer encoder/decoder parameters
$\{(\phi_n,\theta_n)\}_{n=1}^N$ and (ii) the MAC-layer scheduling
policy~$\pi\in\Pi$ to maximize the long-term discounted semantic
significance accrued across the network:
\begin{subequations}
\label{eq:joint_problem_global}
\begin{align}
\max_{\pi \in \Pi,\{\phi_n,\theta_n\}_{n=1}^N} \quad &
\mathbb{E}\!\left[
\sum_{t=0}^{\infty}\gamma^t
\sum_{n=1}^{N}
v_{L_n,Y_{n,t}}\!\left(S_{n,t}\right)
\right]
\label{eq:global_obj} \\
\text{s.t.} \quad
& I(X_{n,t};\mathbf{Z}_{n,t}) \le \eta_{n,t}\,C(\zeta_{n,t}),
  \quad\forall n,t,
\label{eq:global_const_cap} \\
& \sum_{n=1}^N \eta_{n,t} \le W, \quad\forall t,
\label{eq:global_const_mac} \\
& \eta_{n,t} \in \{0,m,2m,\ldots,Km\}, \quad\forall n,t,
\label{eq:global_const_disc}
\end{align}
\end{subequations}
where $\gamma\in[0,1)$ is the discount factor,
$I(X_{n,t};\mathbf{Z}_{n,t})$ is the Shannon mutual information between
the source message~$X_{n,t}$ and the compressed latent~$\mathbf{Z}_{n,t}$, quantifying the source information rate
encoded at the transmitter, $C(\zeta_{n,t})$ is the
Shannon capacity per symbol in~\eqref{eq:capacity}. As defined
in Section \ref{sec:data_significance}, 
$v_{L_n,Y_{n,t}}(s)=D_{L_n}(P_{Y_{n,t}|S_{n,t}=s}\|P_{Y_{n,t}})$ is the
semantic significance of the time-stamped latent $S_{n,t}=s$ for
estimating the target~$Y_{n,t}$ under loss~$L_n$.

Problem \eqref{eq:joint_problem_global} motivates a natural performance metric: \emph{semantic spectrum efficiency}, defined as the \emph{semantic significance delivered per unit bandwidth per unit time (i.e., semantic value per Hz per second)}. Unlike classical spectrum efficiency, which measures spectral
utilization in terms of transmitted bits, semantic spectrum efficiency
quantifies how effectively the available spectrum is used to deliver
task-relevant information for downstream inference. 

\section{\!Significance-Driven Semantic Communication:
A Cross-Layer Design}\label{sec:cross-layer}

\subsection{Problem Decomposition}
Problem~\eqref{eq:joint_problem_global} jointly optimizes (i) semantic encoding and inference
at the physical layer, and (ii) codelength allocation at the MAC layer. The coupling arises through the codelength process
$\{\eta_{n,t}\}$, which
simultaneously determines the channel symbol budget available to
each sensor and the AoI~$\Delta_n(t)$ in the scheduling state.

We decompose problem \eqref{eq:joint_problem_global} via a layered structure. For sensor~$n$, the
encoder-decoder pair is designed to maximize the expected semantic significance that $\hat{\mathbf{Z}}_{n,t}$
 delivers to the receiver:
\begin{subequations}
\label{eq:phy_problem}
\begin{align}
    (\phi_n^*,\theta_n^*) \triangleq \arg\max_{\phi_n,\theta_n}
    \quad &
    \sum_{\delta=1}^{\delta_{\max}}\mathbb{E}_{(\phi_n,\theta_n)}\!\left[v_{L_n,Y_{n,t+\delta}}(\hat{\mathbf{Z}}_{n,t})\right]
    \label{eq:phy_obj}\\
    \text{s.t.} \quad &
    I(X_{n,t};\mathbf{Z}_{n,t}) \le \eta_{n,t}\,C(\zeta_{n,t}).
    \label{eq:phy_const}
\end{align}
\end{subequations}
where $v_{L_n, Y_{n,t+\delta}}(\hat{\mathbf{Z}}_{n,t})$ 
is the semantic significance of the received codeword $\hat{\mathbf{Z}}_{n,t}$ 
for estimating the target $Y_{n,t+\delta}$, and $\delta_{\max}$ is the maximum 
AoI over which the decoder is designed to remain useful. By summing over all possible AoI values $\delta \in \{1, \ldots, 
\delta_{\max}\}$, the encoder-decoder design adapts naturally to the time-varying AoI
$\Delta_n(t)$. Since $\zeta_{n,t}$ and $\eta_{n,t}$ vary dynamically across time slots, problem~\eqref{eq:phy_problem} must be solved for each  realization of these quantities without retraining. A naive implementation would therefore require a distinct encoder-decoder pair per realization, which is computationally prohibitive in a real-time edge network. This challenge is addressed in Section~\ref{sec:meta_vib} through the proposed \emph{Meta-VIB} design.

Let $
g_n(\omega_{n,t},\eta_{n,t}) \triangleq
\mathbb{E}_{(\phi_n^*,\theta_n^*)}\!\left[v_{L_n}(S_{n,t})\mid
\omega_{n,t},\eta_{n,t}\right]$
denote the expected semantic significance delivered by sensor~$n$
under scheduling state
$\omega_{n,t}$ and codelength~$\eta_{n,t}$, when the encoder-decoder pair operates at the optimizer $(\phi_n^*,\theta_n^*)$ of \eqref{eq:phy_problem}. Substituting $g_n$
into \eqref{eq:joint_problem_global} yields the
following MAC-layer scheduling problem:
\begin{subequations}
\label{eq:mac_problem1}
\begin{align}
    \max_{\pi\in\Pi} \quad &
    \mathbb{E}_\pi\!\left[\sum_{t=0}^\infty \gamma^t
    \sum_{n=1}^N g_n\!\left(\omega_{n,t},\eta_{n,t}\right)
    \right] \\
    \text{s.t.} \quad &
    \sum_{n=1}^N \eta_{n,t} \le W, \quad\forall\,t,
    \label{eq:global_const_mac2} \\
    & \eta_{n,t} \in \{0,m,2m,\ldots,Km\}, \quad\forall\,n,t.
\end{align}
\end{subequations}
Problem~\eqref{eq:mac_problem1} is a Multi-Action, Restless Multi-Armed Bandit (MA-RMAB) over a high-dimensional, continuous state space, which poses two key challenges. First, the multi-discrete action space violates the binary-action assumption and indexability required by the classical Whittle-index policy~\cite{whittle1988restless}. Second, the scheduling state $\omega_{n,t}$ is high-dimensional and data-driven, with unknown transition dynamics, causing linear and dynamic programming-based algorithms for finite state space ~\cite{gast2024linear,shisher2023learning,chakraborty2025timely,chamoun2025edge,chen2023index} to suffer from the curse of dimensionality and become impractical. These
two challenges are addressed in Section~\ref{sec:mac_ppo} through the \emph{$Q$-Maximization} scheduling algorithm that was proposed recently in 
\cite{chakraborty2026advantage}.

\subsection{Physical-Layer Design: Meta-VIB}
\label{sec:meta_vib}

We propose \emph{Meta-VIB}, a new \emph{Meta-learning based Variational Information Bottleneck} transceiver design that adapts to time-varying $\zeta_{n,t}$ and $\eta_{n,t}$ without online retraining, thus addressing the physical-layer challenge described after problem \eqref{eq:phy_problem}. 

\begin{figure*}[!htbp]
  \centering
  \includegraphics[width=0.95\textwidth]{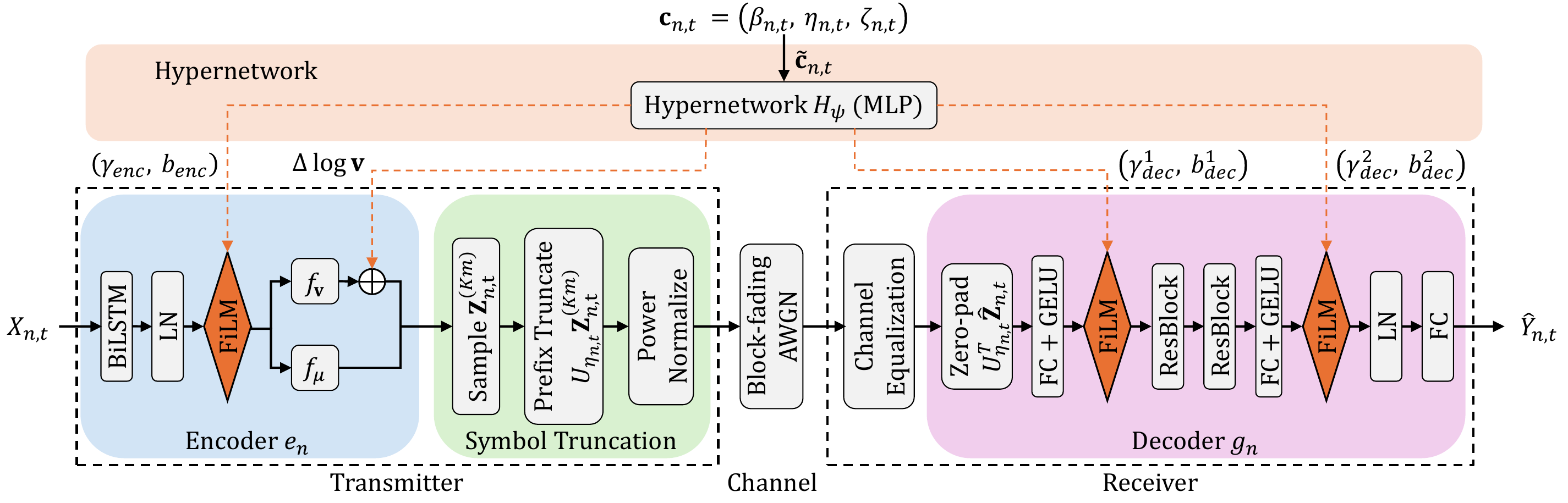}
  \vspace{-2mm}
  \caption{The Meta-VIB encoder-decoder architecture.
  \ignore{
  The hypernetwork
  $H_{\psi_n}$ maps operating point~$c_{n,t}$ to FiLM parameters for the
  encoder and~$L$ decoder stages and to the variance
shift~$\Delta\log\mathbf{v}$. After prefix truncation and power
  normalization, the codeword is transmitted over a block-fading
  channel; after zero-padding, the decoder applies stage-wise FiLM
  conditioning to produce~$\hat{Y}_{n,t}$. }
}\label{fig:phy_hypervib_block}
\vspace{-2mm}
\end{figure*}

\subsubsection{Meta-VIB Formulation}
By Proposition~2.1, problem   \eqref{eq:phy_problem} can be rewritten as a loss-based information bottleneck problem:
\begin{subequations}\label{eq:IB}
\begin{align}
    \max_{\phi_n,\theta_n} \quad & \sum_{\delta=1}^{\delta_{\max}} 
I_{L_n}\!\left(Y_{n,t+\delta};\hat{\mathbf{Z}}_{n,t}\right)
    \label{eq:phy_vib_obj}\\
    \text{s.t.} \quad &
I(X_{n,t};\mathbf{Z}_{n,t}) \le \eta_{n,t}\,C(\zeta_{n,t}).
    \label{relax_vib}
\end{align}
\end{subequations}
Introducing a Lagrange multiplier $\beta_{n,t}\ge 0$
for constraint~\eqref{relax_vib} yields the dual function
\begin{align}
     &d(\beta_{n,t};\eta_{n,t},\zeta_{n,t}) \nonumber\\
\!\triangleq&   \max_{\phi_n,\theta_n}
\!\sum_{\delta=1}^{\delta_{\max}}
    I_{L_n}(Y_{n,t+\delta};\hat{\mathbf{Z}}_{n,t})
    \!-\! \beta_{n,t}\!\Bigl[I(X_{n,t};\mathbf{Z}_{n,t})\! -\! \eta_{n,t}C(\zeta_{n,t})\!\Bigr].
  \label{eq:VIB_Lag}
\end{align}

Both information-theoretic terms on the right-hand side of~\eqref{eq:VIB_Lag} are generally intractable. In the
first term, for each $\delta \in \{1,\ldots,\delta_{\max}\}$,
\begin{align}\label{eq:first_term_bound}
\!\!\!\! &I_{L_n}(Y_{n,t+\delta};\hat{\mathbf{Z}}_{n,t})
  = H_{L_n}(Y_{n,t+\delta}) - H_{L_n}(Y_{n,t+\delta}\mid\hat{\mathbf{Z}}_{n,t}) \nonumber\\
  &\ge\! H_{L_n}(Y_{n,t+\delta})
     - \mathbb{E}\!\left[L_n\!\left(Y_{n,t+\delta},\,g_n(\hat{\mathbf{Z}}_{n,t},\delta;\theta_n)
         \right)
       \right],\!\!
\end{align}
where the inequality follows from  
\begin{align}
    H_{L_n}(Y_{n,t+\delta}|\hat{\mathbf{Z}}_{n,t}) 
    = \inf_{\theta_n}
\mathbb{E}\!\left[L_n\!\left(Y_{n,t+\delta},\,g_n(\hat{\mathbf{Z}}_{n,t},\delta;\theta_n)\right)\right]. 
\end{align}
Since $H_{L_n}(Y_{n,t+\delta})$ is independent of trainable 
parameters, maximizing the first term of \eqref{eq:VIB_Lag} reduces to 
minimizing $\sum_{\delta=1}^{\delta_{\max}} 
\mathbb{E}[L_n(Y_{n,t+\delta},$ $ g_n(\hat{\mathbf{Z}}_{n,t},\delta; \theta_n))]$. 
This replaces the KL-based relaxation used in standard 
VIB~\cite{alemi2016deep} with the $L_n$-conditional-entropy  
relaxation in \eqref{eq:first_term_bound}, thereby aligning the Meta-VIB objective directly with 
the task loss across all AoI values.
  
For the second term, let 
$q_{\phi_n}(\mathbf{Z}_{n,t}\mid X_{n,t})$ denote the encoder’s 
posterior. Following \cite{alemi2016deep}, we introduce
the variational prior
$r(\mathbf{Z}_{n,t})=\mathcal{CN}(\mathbf{0},\sigma_r^{2}\mathbf{I}_{\eta_{n,t}})$. By the nonnegativity of the KL divergence, one can obtain  \cite{alemi2016deep}
\begin{align}
  I(X_{n,t};\mathbf{Z}_{n,t})
  \le \mathbb{E}_{X_{n,t}}\!\left[
        D_{\mathrm{KL}}\!\left(
          q_{\phi_n}(\mathbf{Z}_{n,t}\mid X_{n,t})
          \;\|\; r(\mathbf{Z}_{n,t})
        \right)
      \right].
  \label{eq:upper_bound}
\end{align}
To reduce mismatch between the aggregated posterior
$q_{\phi_n}(\mathbf{Z}_{n,t})=\mathbb{E}[q_{\phi_n}(\mathbf{Z}_{n,t}\mid X_{n,t})]$ and the Gaussian prior $r(\mathbf{Z}_{n,t})$, we further
include a Maximum Mean Discrepancy (MMD) penalty~\cite{gretton2012kernel}.
Substituting~\eqref{eq:first_term_bound} and~\eqref{eq:upper_bound}
into~\eqref{eq:VIB_Lag},  adding the MMD term, and removing the non-trainable term, yields the Meta-VIB objective
\begin{align}
\mathcal{L}_{\mathrm{VIB}}
=
\min_{\phi_n,\theta_n}\;
\mathbb{E}\Bigg[
&\sum_{\delta=1}^{\delta_{\max}}
L_n\!\left(
Y_{n,t+\delta},
g_n(\hat{\mathbf Z}_{n,t},\delta;\theta_n)
\right)
\notag\\
&\quad+
\beta_{n,t}\,
D_{\mathrm{KL}}\!\left(
q_{\phi_n}(\mathbf Z_{n,t}\mid X_{n,t})
\,\|\, r(\mathbf Z_{n,t})
\right)
\Bigg]
\notag\\
&\quad+
\beta_{n,t}\,
\mathrm{MMD}^2\!\left(
q_{\phi_n}(\mathbf Z_{n,t}),
r(\mathbf Z_{n,t})
\right).
\label{eq:vib_objective}
\end{align}


\subsubsection{FiLM-based Lightweight Meta-VIB Design.}
\label{FiLM_meta}

The key novelty of Meta-VIB is
a single neural architecture that generalizes across all realizations of $\zeta_{n,t}$, $\eta_{n,t}$, and $\beta_{n,t}$ without retraining. 
Unlike VIB~\cite{alemi2016deep}, which requires a separately optimized model for each $\beta_{n,t}$, and Hyper-VIB~\cite{peng2025hyper}, which adapts only to $\beta_{n,t}$, Meta-VIB generalizes jointly across all three variables via Feature-wise Linear Modulation (FiLM)~\cite{perez2018film}.
As illustrated in Fig.~\ref{fig:phy_hypervib_block}, the FiLM layers scale and shift the intermediate features of the encoder and decoder, conditioned on the operating point
$c_{n,t} \triangleq (\zeta_{n,t}, \eta_{n,t}, \beta_{n,t})$. 
FiLM's parameter count scales linearly with the feature dimension, making it significantly more efficient than attention~\cite{vaswani2017attention}, which is unnecessarily expensive for this low-dimensional operating point. 
To the best of our knowledge, this is the first use of FiLM in semantic communication.

\paragraph*{Encoder and Decoder Backbone.} 
The encoder, parameterized by $\phi_n$, uses a BiLSTM~\cite{schuster1997bidirectional} to map the historical feature sequence $X_{n,t}$ to a hidden representation
$\mathbf{h}_{\mathrm{raw},n,t} = f_{\mathrm{enc}}(X_{n,t};\phi_n)$. The decoder, parameterized by $\theta_n$, consists of $R$ residual Multi-Layer Perceptron (MLP) blocks with Gaussian Error Linear Unit (GELU) activations. 
\begin{figure*}[!htbp]
  \centering
  \includegraphics[trim={0 21.4cm 0 0}, clip,
    width=0.9\textwidth]{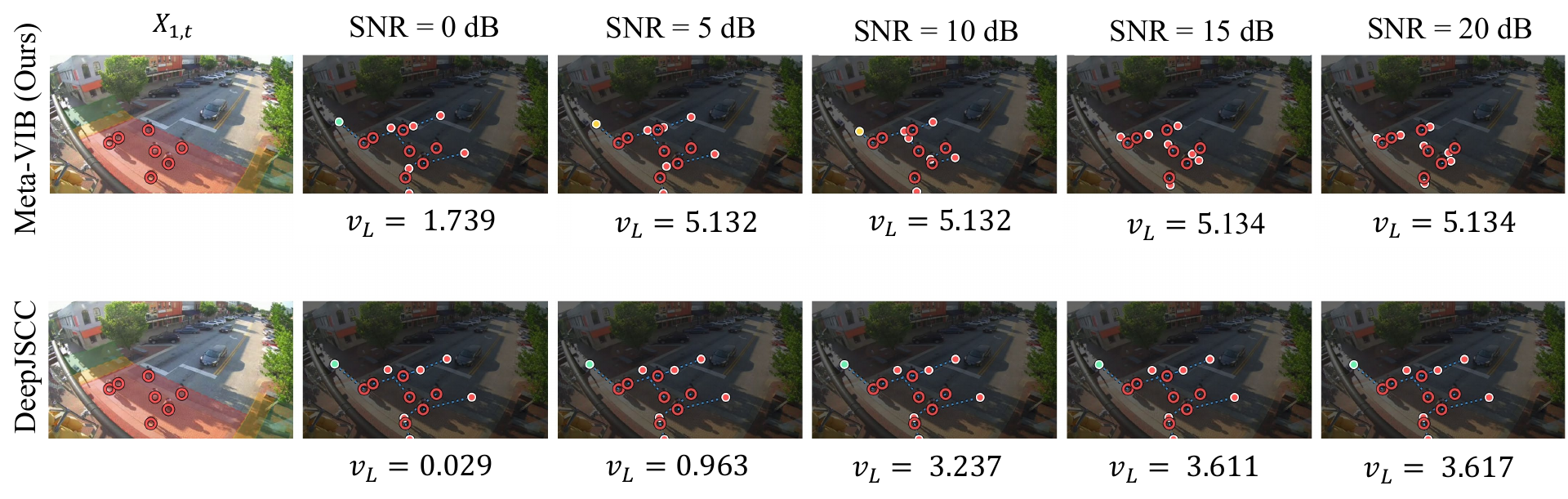}
    \caption{Per-sample semantic significance $v_{L_n,Y_{n,t+\delta}}(z)$ (denoted simply as $v_L$ in the figure) versus the
    SNR $\zeta_{n,t}$, where the codelength is $\eta_{n,t}=2$ and the AoI is $\delta=30$. 
    }
  \label{fig:visual_comparison}
  \vspace{-2mm}
\end{figure*}
\paragraph*{FiLM Conditioning.}
The operating point $c_{n,t}$ is fed to a 
lightweight MLP hypernetwork $H_{\psi_n}$, parameterized by $\psi_n$, which modulates the encoder and decoder backbone through FiLM layers~\cite{perez2018film} without changing the backbone weights. In a single forward pass, $H_{\psi_n}$ outputs all modulation parameters:
\begin{align}
    \bigl(\gamma_{\mathrm{enc}},\, b_{\mathrm{enc}},\,
    \{\gamma_{\mathrm{dec}}^{{r}},\, b_{\mathrm{dec}}^{{ r}}\}_{{ r}=1}^{{ R}},\,
    \Delta\log\mathbf{v}\bigr)
    = H_{\psi_n}(c_{n,t}).
\end{align}
FiLM is applied once in the encoder, just before the bottleneck:
\begin{align}
    \mathbf{h}_{n,t}
    = \gamma_{\mathrm{enc}} \odot \mathrm{LN}(\mathbf{h}_{\mathrm{raw},{n,t}})
      + b_{\mathrm{enc}},
\end{align}
where $\mathrm{LN}(\cdot)$ denotes Layer Normalization and $\odot$ denotes element-wise multiplication. From $\mathbf{h}_{n,t}$, the encoder produces the posterior parameters $\boldsymbol{\mu}_{n,t}\in\mathbb{C}^{Km}$ and $\log\mathbf{v}_{n,t}\in\mathbb{R}^{Km}$:
\begin{align}
    \boldsymbol{\mu}_{n,t} &= f_{\mu}(\mathbf{h}_{{n,t}};\phi_n), \\
    \log\mathbf{v}_{n,t}   &= f_{v}(\mathbf{h}_{{n,t}};\phi_n)
                               + \Delta\log\mathbf{v}.
\end{align}
These define the posterior $q_{\phi_n}(\mathbf{Z}_{n,t}^{(Km)}\mid X_{n,t})
=\mathcal{CN}(\boldsymbol{\mu}_{n,t},\mathrm{diag}(\mathbf{v}_{n,t}))$, from which $\mathbf{Z}_{n,t}^{(Km)}\in\mathbb{C}^{Km}$ is sampled via the reparameterization trick.
The KL divergence upper bound on $I(X_{n,t};\mathbf{Z}_{n,t}^{(Km)})$ is \cite{alemi2016deep}
\begin{align}
    \sum_{k=1}^{Km}\!\left[
      \frac{|[\boldsymbol{\mu}_{n,t}]_k|^2}{\sigma_r^2}
      + \frac{[\mathbf{v}_{n,t}]_k}{\sigma_r^2}
      - \log\frac{[\mathbf{v}_{n,t}]_k}{\sigma_r^2} - 1
    \right],
\end{align}
where $[\mathbf{x}]_k$ denotes the $k$th element of vector $\mathbf{x}$. Thus, both the mean and the variance contribute to the information rate; however, the variance $[\mathbf{v}_{n,t}]_k$ is the dominant factor: a smaller variance yields a tighter posterior and therefore preserves more information at coordinate $k$. FiLM is also applied to each of the $R$ decoder blocks, allowing the conditioning signal to shape the entire reconstruction process (see Fig.~\ref{fig:phy_hypervib_block}).

\paragraph*{Latent Truncation, Power Normalization, and Zero-Padding.}
Let $U_{\eta_{n,t}}\in\{0,1\}^{\eta_{n,t}\times Km}$ be the
prefix-selection matrix. The full latent vector $\mathbf{Z}_{n,t}^{{ (Km)}}$ is truncated to its first $\eta_{n,t}$ elements and then power-normalized as
\begin{align}
    \mathbf{Z}_{n,t}
    = \sqrt{\eta_{n,t}P}\;
      \frac{U_{\eta_{n,t}}\mathbf{Z}_{n,t}^{{ (Km)}}}
           {\|U_{\eta_{n,t}}\mathbf{Z}_{n,t}^{{(Km)}}\|_2}
    \;{\in\mathbb{C}^{\eta_{n,t}}},
    \label{eq:power_norm}
\end{align}
which ensures $\mathbb{E}[\|\mathbf{Z}_{n,t}\|_2^2\mid\eta_{n,t}]=\eta_{n,t}P$. After block-fading transmission in~\eqref{Gaussian_Channel}, the transpose $U_{\eta_{n,t}}^{\top}$ maps the received latent $\hat{\mathbf{Z}}_{n,t}$ back to a $Km$-dimensional vector 
$\hat{\mathbf{Z}}_{n,t}^{(Km)}
    = U_{\eta_{n,t}}^{\top}\hat{\mathbf{Z}}_{n,t}.$
\ignore{    
\begin{align}
    \tilde{\mathbf{Z}}_{n,t}
    = U_{\eta_{n,t}}^{\top}\hat{\mathbf{Z}}_{n,t}.
\end{align}}
The decoder then maps the most recently received latent $\hat{\mathbf{Z}}_{n,t-\delta}^{(Km)}$, which arrived $\Delta_n(t)=\delta$ slots ago, along with the current AoI $\delta$, 
to the 
estimate $\hat{Y}_{n,t}$ through $R$ FiLM-conditioned residual blocks.

\paragraph*{Information-Concentrating Regularizer.} 
Latent truncation is effective only when the leading latent coordinates carry more information than the trailing ones. Since a smaller posterior variance at coordinate $k$ indicates that the encoder has concentrated more information there, enforcing the non-decreasing profile
$[\mathbf{v}_{n,t}]_1 \le [\mathbf{v}_{n,t}]_2 \le \cdots \le
[\mathbf{v}_{n,t}]_{Km}$ concentrates information in the leading dimensions and ensures that prefix truncation discards the least informative suffix first. We enforce this structure via a hinge regularizer on the order of the log-variances 
\begin{align}
   \!\!\! L_{\mathrm{order}}(\mathbf{v}_{n,t})
    =& \frac{1}{Km-1}\nonumber\\&\sum_{k=1}^{Km-1}
      \max\bigl(0,\;[\log\mathbf{v}_{n,t}]_{k}
        - [\log\mathbf{v}_{n,t}]_{k+1}
      \bigr),\!
    \label{eq:lord}
\end{align}
which is zero when the log-variances are non-decreasing and positive otherwise. Working in log-space is natural because the encoder is parameterized in terms of log-variance, and the monotonicity of the exponential function ensures that ordering the log-variances also orders the variances themselves. The final training objective augments~\eqref{eq:vib_objective} with the expected hinge regularizer:
\begin{align}
\mathcal{L}_{\mathrm{train}}
=
\mathcal{L}_{\mathrm{VIB}}
+
\rho_{\mathrm{ord}}\,
\mathbb{E}\!\left[
L_{\mathrm{order}}(\mathbf v_{n,t})
\right],
\label{eq:loss}
\end{align}
where $\rho_{\mathrm{ord}}>0$ is a weighting parameter.
\ignore{
Training Meta-VIB end-to-end from scratch over the entire region of $c_{n,t}$ is unstable. 
To address this issue, a carefully designed three-stage training procedure is provided in \ifreport
Appendix~\ref{app:training}.
\else
\cite{mcdowell2026significance}.
\fi 
}

\paragraph*{Meta-VIB Training.}
Training Meta-VIB end-to-end from scratch over the entire region of $c_{n,t}$ is unstable. We  propose a three-phase training procedure to resolve this issue. 
\emph{(i)}~\emph{Joint pretraining:} 
All parameters $\{\phi_n,\theta_n,\psi_n\}$ are first optimized jointly using a fixed Lagrange multiplier $\beta_{n,t}=\beta_0$, where $\beta_0 = 10^{-3}$ in our implementation. Fixing $\beta_{n,t}$ allows the encoder and decoder to first learn a stable and semantically meaningful latent representation before the hypernetwork begins adapting the model to dynamic $\beta_{n,t}$.
\emph{(ii)}~\emph{Hypernetwork training:} the backbone is kept frozen and only $\psi_n$ is updated, with $\beta_{n,t}$ drawn log-uniformly from $[\beta_{\min},\beta_{\max}]$. This forces the hypernetwork $H_{\psi_n}$ to modulate the learned backbone through FiLM, rather than co-adapting with the backbone weights.
\emph{(iii)}~\emph{Joint fine-tuning:} 
all parameters are unfrozen and optimized with a small learning rate (e.g., $1.5\times10^{-5}$ in our case) for the backbone and a larger learning rate (e.g., $7.5\times10^{-5}$) for the hypernetwork, while training cycles through the discrete grid $(\beta_{n,t},\eta_{n,t}) \in \{(\beta_i,\eta_j)\}$ and continues to sample $\zeta_{n,t}$ randomly to reduce residual mismatch. In our implementation, the three phases occupy $40\%$, $45\%$, and $15\%$ of the total training epochs, respectively. Without this three-stage training procedure, Meta-VIB would not perform well.

\paragraph*{Online Dual Optimization.} 
At deployment, for each realization of $(\eta_{n,t}, \zeta_{n,t})$ at slot $t$, the optimal dual variable $\beta_{n,t}^{*}$ is computed online by maximizing the dual function in~\eqref{eq:VIB_Lag} via golden-section search over $\log\beta \in [\log\beta_{\min}, \log\beta_{\max}]$, which converges in 25 iterations.

\subsection{MAC-Layer Design: $Q$-Maximization Algorithm} \label{sec:mac_ppo}
Problem~\eqref{eq:mac_problem1} is a Multi-Action Restless Multi-Armed Bandit (MA-RMAB). To solve it, we apply   Lagrangian relaxation and dual decomposition.
Specifically, the hard per-slot constraint ~\eqref{eq:global_const_mac2} is relaxed to a discounted budget constraint, and a dual variable $\lambda \geq 0$ is introduced. For each fixed $\lambda$, the Lagrangian separates across sensors, and the dual
function decomposes into  $N$ independent per-sensor MDPs with reward
$g_n(\omega_{n,t}, \eta_{n,t}) - \lambda\eta_{n,t}$;  see 
\ifreport
Appendix~\ref{app:advantage} for details.
\else
\cite{mcdowell2026significance} for  details.
\fi 

\begin{figure*}[htbp]
    \centering
    \subfloat[{Varying codelength $\eta_{n,t}$ for SNR $\zeta_{n,t}= 0$ dB and AoI $\delta=100$.}]{
        \includegraphics[width=0.31\textwidth]
          {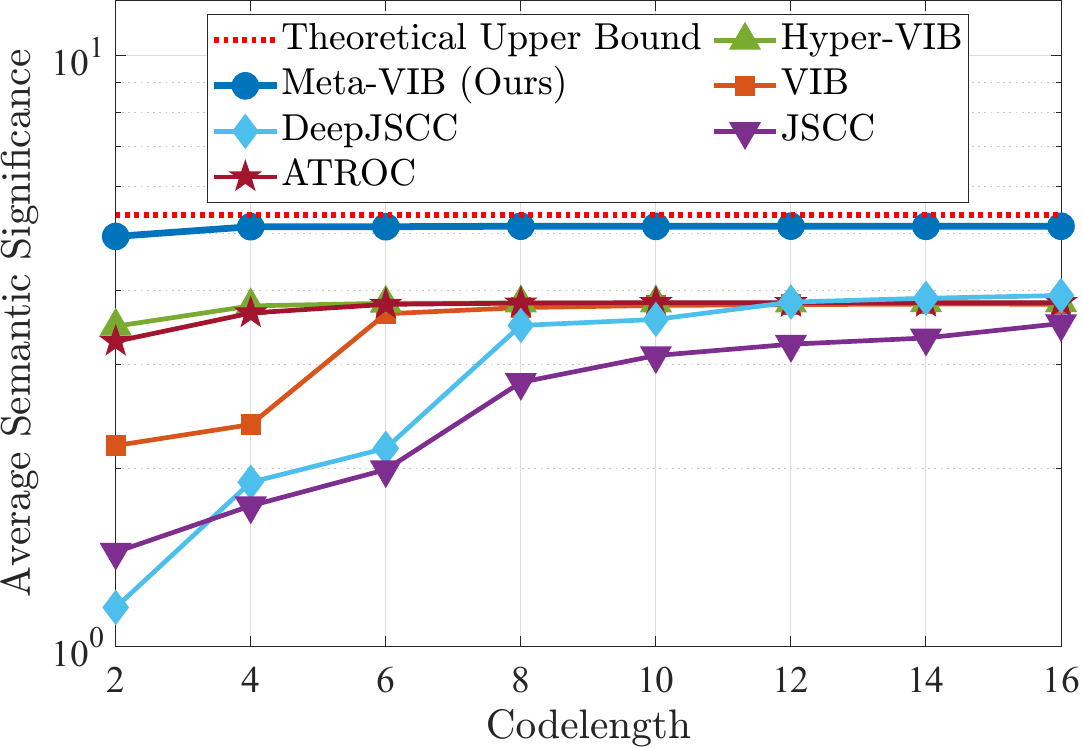}
        \label{fig:sig_vs_eta}}
    \hfill
    \subfloat[{Varying SNR $\zeta_{n,t}$ for codelength $\eta_{n,t}=2$ and AoI $\delta=100$.}]{
        \includegraphics[width=0.30\textwidth]{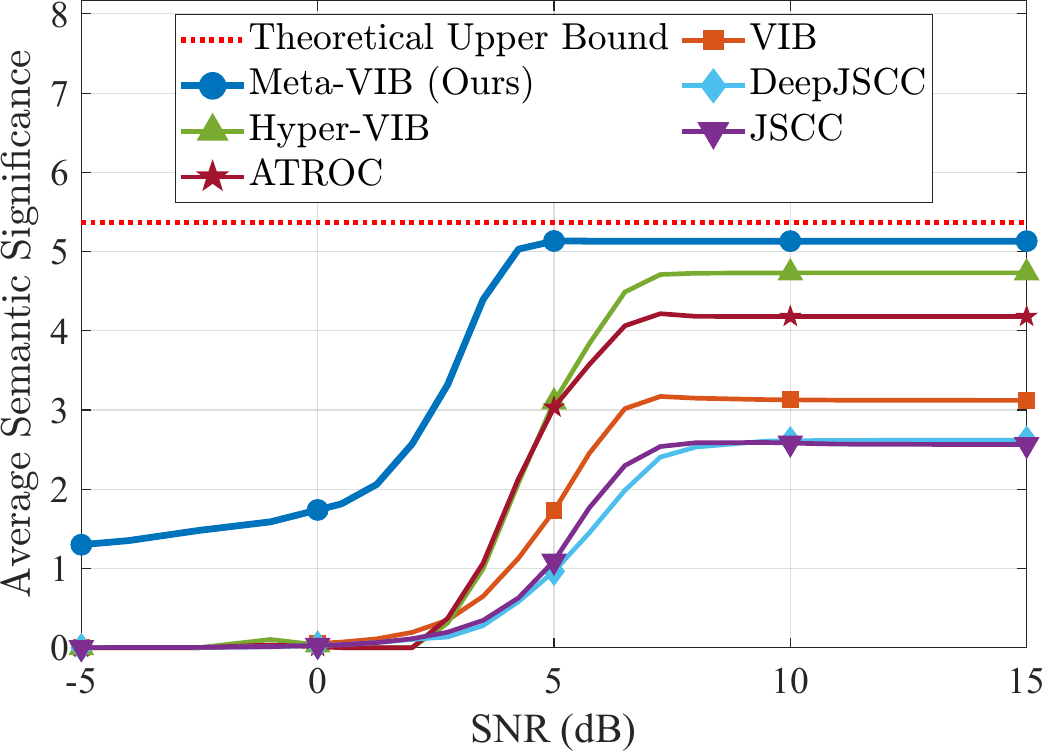}
        \label{fig:sig_vs_snr}}
    \hfill
    \subfloat[{Varying AoI $\delta$ for codelength $\eta_{n,t}=2$ and SNR $\zeta_{n,t}=0$ dB.}]{
        \includegraphics[width=0.32\textwidth]
          {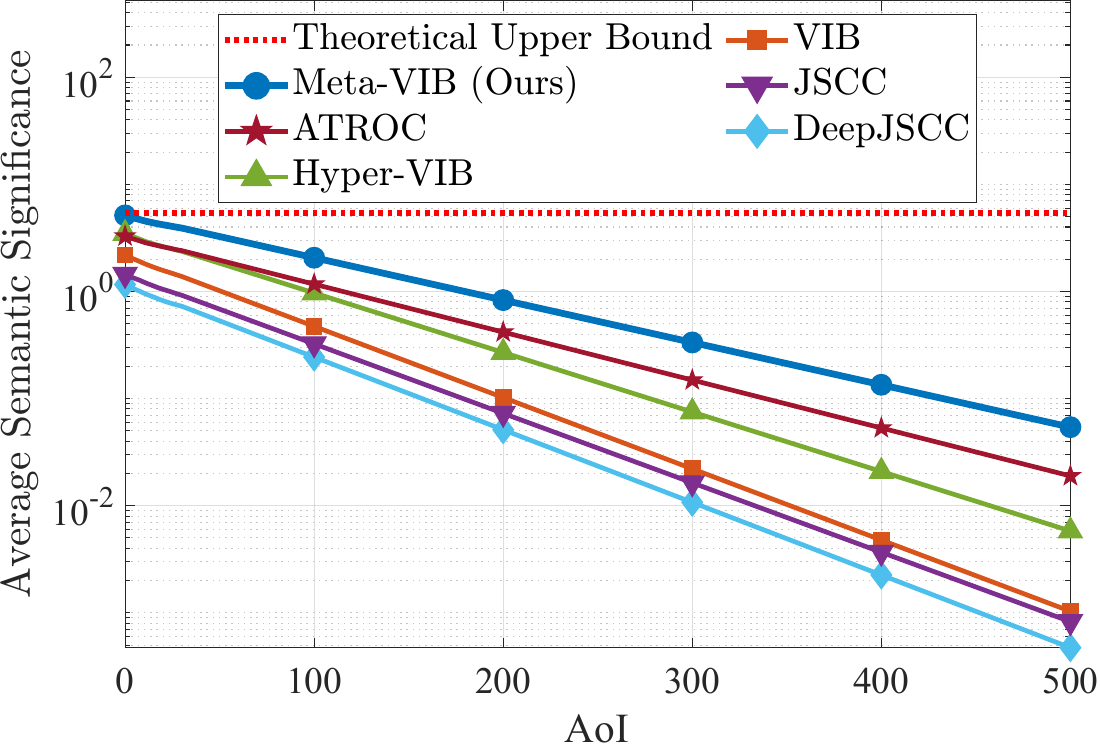}
        \label{fig:sig_vs_aoi}}
    \caption{Average semantic significance $\bar{v}_{L_n,Y_{n,t+\delta},\hat{\mathbf{Z}}_{n,t}}$ versus the codelength $\eta_{n,t}$, the instantaneous  SNR $\zeta_{n,t}$, and the AoI $\delta$ for different physical-layer designs.} 
    \label{fig:sig_per_bit_comparison}
    
\end{figure*}

\begin{algorithm}[t]
\caption{$\lambda$-Conditioned  Actor-Critic Algorithm 
}
\label{alg:kam_ppo}
\begin{algorithmic}[1]
\State Initialize actor-critic network with policy head $\pi_{\varphi_n}(\eta \mid \tilde{\omega})$, value head $V_{\varphi_n}(\tilde{\omega})$, and $Q$-head $Q_{\varphi_n}(\tilde{\omega}, \eta)$;
\For{each training iteration}
    \State Sample $\lambda \sim \mathrm{Uniform}([0, \lambda_{\max}])$;
    \For{each slot $t$ in rollout}
        \State Observe $\omega_{n,t}$ and form $\tilde{\omega}_{n,t} = (\omega_{n,t}, \lambda)$;
        \State Select $\eta_{n,t} \sim \pi_{\varphi_n}(\cdot \mid \tilde{\omega}_{n,t})$; 
        \State Receive reward $g_n(\omega_{n,t}, \eta_{n,t}) - \lambda\,\eta_{n,t}$;
    \EndFor
    \State Update $\pi_{\varphi_n}$, $V_{\varphi_n}$, and $Q_{\varphi_n}$ via PPO;
\EndFor
\end{algorithmic}
\end{algorithm}

\subsubsection{$\lambda$-Conditioned  Actor-Critic Offline Training}\label{sec:lambda_rl}
The optimal dual variable $\lambda^*$ depends on system configuration, such as the number of sensors $N$ and the available bandwidth $W$. Existing RL methods for restless bandits~\cite{nakhleh2021neurwin, killian2021q, avrachenkov2020whittle, xiong2022index} typically assume finite state spaces and require retraining when such configurations change. To overcome this, we learn per-sensor $Q$-functions offline over a continuous range of dual variables $\lambda$, so that at deployment, adaptation to any given configuration requires only computing $\lambda^*$, which is substantially cheaper than retraining the RL model.

Following~\cite{nakhleh2022deeptop},
we augment the per-sensor state as $\tilde{\omega}_{n,t} \triangleq (\omega_{n,t}, \lambda)$
and sample $\lambda\sim\text{Uniform}([0, \lambda_{\max}])$ once per rollout, so that a single network learns the $Q$-function as a continuous function of $\lambda$. We train a shared actor-critic network with three heads \cite{haarnoja2018soft}: (i) a policy head
$\pi_{\varphi_n}(\eta_{n,t} \mid \tilde{\omega}_{n,t})$, (ii) a value head $V_{\varphi_n}(\tilde{\omega}_{n,t})$, and (iii) a
per-action $Q$-head $Q_{\varphi_n}(\tilde{\omega}_{n,t}, \eta_{n,t})$. Using  PPO~\cite{schulman2017proximal}, the network is trained directly from data over the high-dimensional, continuous state space. At deployment,
the per-sensor Q-values $Q_{\varphi_n}(\tilde{\omega}_{n,t}, \eta_{n,t})$ are passed to the online
scheduler. The full offline training procedure is summarized in Algorithm~\ref{alg:kam_ppo}.

\subsubsection{
Online Scheduling}
At deployment, we first compute the optimal dual variable $\lambda^*$ for the current system configuration by bisection; see \ifreport
Appendix~\ref{app:advantage} for details.
\else
\cite{mcdowell2026significance} for  details.
\fi We then evaluate the learned per-sensor Q-values $Q_{\varphi_n}(({\omega}_{n,t},\lambda^*),\eta_{n,t})$ using $\lambda^*$ and determine the codelength allocation at each slot $t$ by using the Q-Maximization algorithm \cite{chakraborty2026advantage} to solve the following Multiple-Choice Knapsack Problem (MCKP):   
\begin{subequations}\label{eq:mckp}
\begin{align}
  \max_{\{\eta_{n,t}\}} \quad
  & \sum_{n=1}^{N}
    Q_{\varphi_n}\!\left(\left(\omega_{n,t}, \lambda^*\right),
    \eta_{n,t}\right)
    \label{eq:mckp_obj} \\
  \text{s.t.} \quad
  & \sum_{n=1}^{N} \eta_{n,t} \leq W,
    \label{eq:mckp_budget} \\
  & \eta_{n,t} \in \{0, m, 2m, \ldots, Km\}, \quad \forall~ n,
    \label{eq:mckp_action}
\end{align}
\end{subequations}
where 
\eqref{eq:mckp_budget} enforces the global spectrum budget.

Traditionally, MCKP scheduling problems, such as~\eqref{eq:mckp}, are
solved via dynamic programming~\cite{shisher2023learning}
with a per-slot complexity $O(NWK)$ \cite{kellerer2004multidimensional}, where $N$~is the number of
sensors, $W$~is the shared symbol budget, and $K$~is the number of
codelength levels per sensor. Nonetheless, it was recently shown
in~\cite{chakraborty2026advantage} that solving \eqref{eq:mckp} via standard dynamic programming is
not asymptotically optimal for problem~\eqref{eq:mac_problem1} when $N$ and $W$ grow proportionally to infinity.
To
address this limitation, \cite{chakraborty2026advantage}~introduced a
class of multi-action Linear Program (LP)-priority policies and proved their
asymptotic optimality for MA-RMABs with finite state and action spaces, generalizing Verloop's LP-priority policies for binary-action RMABs~\cite{verloop2016asymptotically} to the more challenging MA-RMABs.
The $Q$-Maximization algorithm belongs to this class and is therefore asymptotically optimal for finite state and action spaces. To enforce the LP-priority rules, $Q$-Maximization augments the dynamic programming procedure with an LP-priority-based tie-breaking step, and the overall complexity is still $O(NWK)$. Due to space limitations, the detailed steps
of  $Q$-Maximization are provided in
\ifreport Appendix~\ref{app:lp_priority}.
\else 
\cite{mcdowell2026significance}. 
\fi 
Moreover, to handle continuous state spaces, $Q$-Maximization employs the $\lambda$-conditioned actor-critic algorithm described in Section~\ref{sec:lambda_rl} to compute the Q-values  $Q_{\varphi_n}(({\omega}_{n,t},\lambda ),\eta_{n,t})$. 
Fig.~\ref{fig:mac_evaluation} compares
$Q$-Maximization against 
the Net-Gain Maximization (NGM) policy \cite{shisher2023learning}, which applies dynamic programming without LP-priority based tie-breaking.

\ignore{
\subsection{MAC-Layer Design: $Q$-Maximization Algorithm} \label{sec:mac_ppo}

Problem~\eqref{eq:mac_problem1} is a Multi-Action Restless Multi-Armed Bandit (MA-RMAB). To solve it, we adopt the low-complexity \emph{$Q$-Maximization} algorithm in~\cite{chakraborty2026advantage}, which allocates channel resources dynamically according to the per-sensor $Q$-functions.

We introduce a dual multiplier $\lambda \ge 0$ to relax the  constraint~\eqref{eq:global_const_mac2}. By  dual decomposition, the problem \eqref{eq:mac_problem1} separates into $N$ independent per-sensor MDPs with a  $\lambda$-penalized reward $g_n(\omega,\eta) - \lambda\,\eta$; see 
\ifreport
Appendix~\ref{app:advantage} for the details.
\else
\cite{mcdowell2026significance} for the details.
\fi

We design an RL algorithm that is adaptive to varying number of sensors~$N$ and available bandwidth~$W$ without requiring for retraining. Changes in the the number of sensors~$N$ and available bandwidth~$W$ are captured by the optimal dual multiplier~$\lambda^*$. To adapt to these variations, we augment the per-sensor state as $\tilde{\omega}\triangleq(\omega_{n,t},\lambda)$ following~\cite{nakhleh2022deeptop}, and sample $\lambda$ uniformly from $[0,\lambda_{\max}]$ across rollouts.

We then train a single actor-critic network with three heads---a policy head $\pi_\theta(\eta \mid \tilde{\omega})$, a value head $V_\theta(\tilde{\omega})$, and a per-action Q-head---via PPO~\cite{schulman2017proximal} to learn
\begin{align}
  Q_{\theta}(\tilde{\omega},\eta)
  \approx
  g_n(\omega,\eta) - \lambda\,\eta
  + \gamma\,\mathbb{E}\!\left[V_{\theta}(\tilde{\omega}') \mid \tilde{\omega},\eta\right]
  \label{eq:q_def}
\end{align}
directly from data over the high-dimensional, continuous state space. At deployment, the per-sensor scores $Q_\theta(\tilde{\omega}_{n,t}, \eta)$ are passed to the online scheduler. The full training procedure is summarized in Algorithm~\ref{alg:kam_ppo}.


This allows the network to learn $Q_\theta(\tilde{\omega}, \eta)$ over a continuous range of dual prices, so that a single trained model adapts to new system configurations (e.g., available bandwidth or number of sensors) by recovering the appropriate $\lambda^{*}$ via bisection without retraining. 

To learn the $Q$-values in \eqref{eq:q_def} directly from data, we train a single actor-critic network with three heads: a policy head $\pi_\theta(\eta \mid \tilde{\omega})$, a value head $V_\theta(\tilde{\omega})$, and a per-action $Q$-head $Q_\theta(\tilde{\omega}, \eta)$, using  PPO~\cite{schulman2017proximal}. The training procedure is summarized in Algorithm~\ref{alg:kam_ppo}.

\begin{algorithm}[t]
\caption{$\lambda$-Conditioned Actor-Critic Training for the $Q$-Maximization algorithm}
\label{alg:kam_ppo}
\begin{algorithmic}[1]
\State Initialize network with policy head $\pi_\theta(\eta \mid \tilde{\omega})$, value head $V_\theta(\tilde{\omega})$, and $Q$-head $Q_\theta(\tilde{\omega}, \eta)$
\For{each training iteration}
    \State Sample $\lambda \sim \mathrm{Uniform}[0, \lambda_{\max}]$
    \For{each slot in rollout}
        \State Observe $\omega_{n,t}$; form $\tilde{\omega} = (\omega_{n,t}, \lambda)$
        \State Select $\eta \sim \pi_\theta(\cdot \mid \tilde{\omega})$; receive reward $g_n(\omega_{n,t}, \eta) - \lambda\,\eta$
    \EndFor
    \State Update $\pi_\theta$, $V_\theta$, and $Q_\theta$ via PPO
\EndFor
\end{algorithmic}
\end{algorithm}

\subsubsection{$Q$-Maximization Algorithm for Online Scheduling}
At deployment, the per-sensor $Q$-functions from the trained network are used to select codelength allocations at each slot~$t$ by solving
\begin{subequations}\label{eq:mckp}
\begin{align}
\max_{\{B_{n,\eta}(t)\}} \quad &
\sum_{n=1}^{N}\sum_{\eta\in\mathcal{A}}
Q_{\theta}(\tilde{\omega}_{n,t},\eta)\,B_{n,\eta}(t)
\label{eq:mckp_obj}\\
\text{s.t.}\quad &
\sum_{n=1}^{N}\sum_{\eta\in\mathcal{A}}\eta\,B_{n,\eta}(t) \le W,
\label{eq:mckp_budget}\\
&
\sum_{\eta\in\mathcal{A}}B_{n,\eta}(t) = 1,\quad \forall n,
\label{eq:mckp_single}\\
&
B_{n,\eta}(t) \in \{0,1\},\quad \forall n,\eta,
\label{eq:mckp_binary}
\end{align}
\end{subequations}
where $B_{n,\eta}(t)\in\{0,1\}$ indicates whether codelength~$\eta$ is assigned to sensor~$n$, constraint~\eqref{eq:mckp_budget} enforces the shared spectrum budget, and~\eqref{eq:mckp_single}--\eqref{eq:mckp_binary} ensure exactly one action per sensor, including $\eta_{n,t}=0$ for no transmission. Problem~\eqref{eq:mckp} is a multiple-choice knapsack problem (MCKP). The $Q$-Maximization algorithm is proven to be asymptotically optimal in~\cite{chakraborty2026advantage}. \cite{chakraborty2026advantage} develops a family of low-complexity scheduling policies for MA-RMABs, among which $Q$-Maximization is employed in this work for its simplicity and direct compatibility with deep RL.
}
\vspace{-3mm}
\section{Data-Driven Evaluation} \label{sec:evaluation}

This section presents data-driven evaluation results of the proposed semantic communication design.

\vspace{-3mm}

\subsection{\!\!\!Data Collection, Loss Function, and Baselines}
\label{sec6.1}
Our evaluation uses live feeds from four synchronized public traffic cameras at Toomer's Corner, a busy downtown intersection in Auburn, Alabama~\cite{toomers2024webcam}. Each camera monitors one crosswalk; together, they cover the central X-shaped pedestrian crossing area. During the all-stop phase, all vehicle traffic is halted and pedestrians may cross in any direction, including diagonally.
A Raspberry Pi~5 captured 1080p RTSP streams at 20 fps from 06:00 to 18:00 over a two-week period (Feb.~17--Mar.~3, 2026). The recorded frames were processed with a pre-trained YOLOv8s + DeepSORT pipeline: YOLOv8s detects pedestrian bounding boxes and traffic-light states (green, yellow, or red), while DeepSORT maintains persistent pedestrian identities across frames. The crossing area is divided into three color-coded risk zones: green (\textit{safe}), yellow (\textit{cautious}), and red (\textit{dangerous}), and each pedestrian bounding box is assigned a safety label based on its occupied zone.
\ifreport
Appendix \ref{sec:appendix_visuals}
presents
\else
In \cite{mcdowell2026significance}, we present
\fi
annotated video frames from all four cameras, showing risk zones, pedestrian locations, and their safety labels.
\begin{figure*}[htbp]
    \centering
    \subfloat[{Varying channel budget $W$ for average SNR $\mathbb{E}[\zeta_{n,t}]=0$ dB and No. of sensors $N=1000$.}]{
        \includegraphics[width=0.31\textwidth]{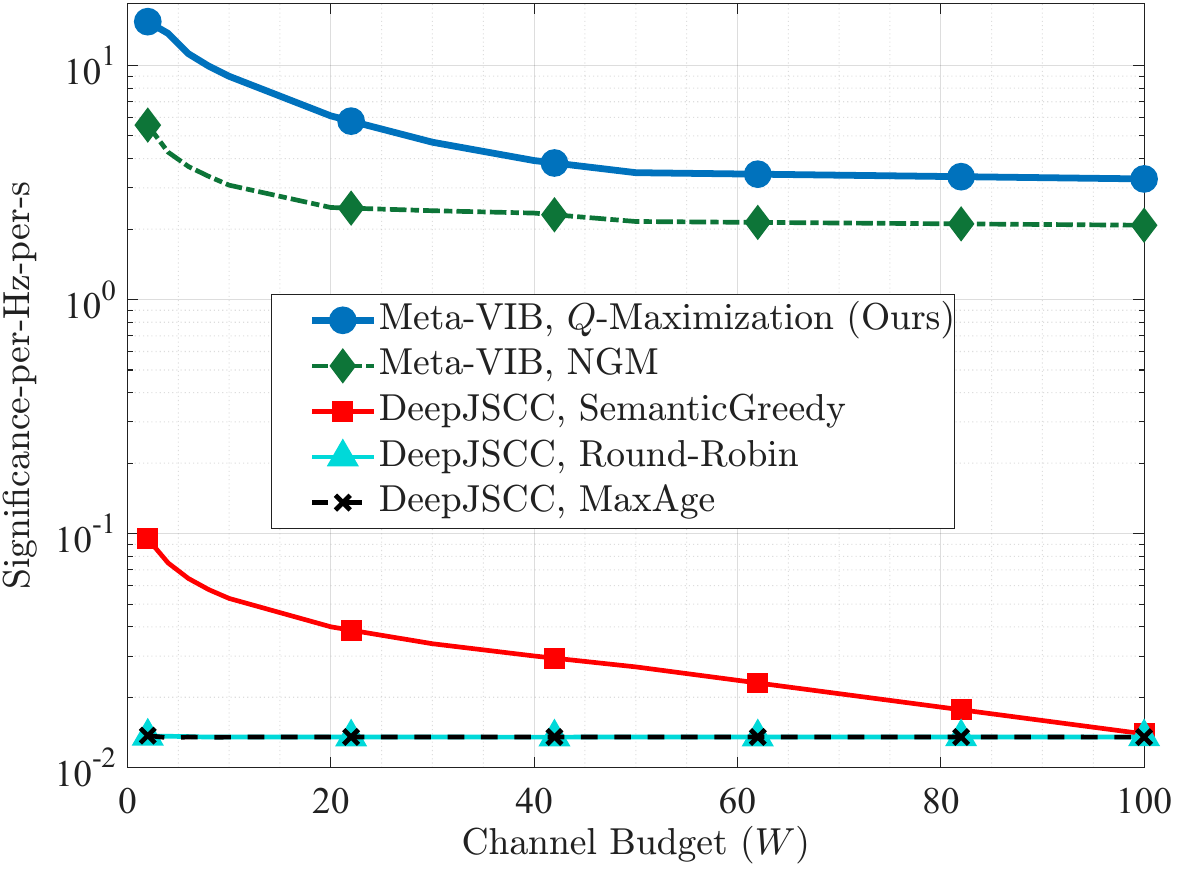}
        \label{fig:sig_per_symbol_budget}}
    \hspace{1mm}
    \subfloat[{Varying average SNR $\mathbb{E}[\zeta_{n,t}]$ for channel budget $W=40$ and No. of sensors $N=1000$.}]{
        \includegraphics[width=0.31\textwidth]
          {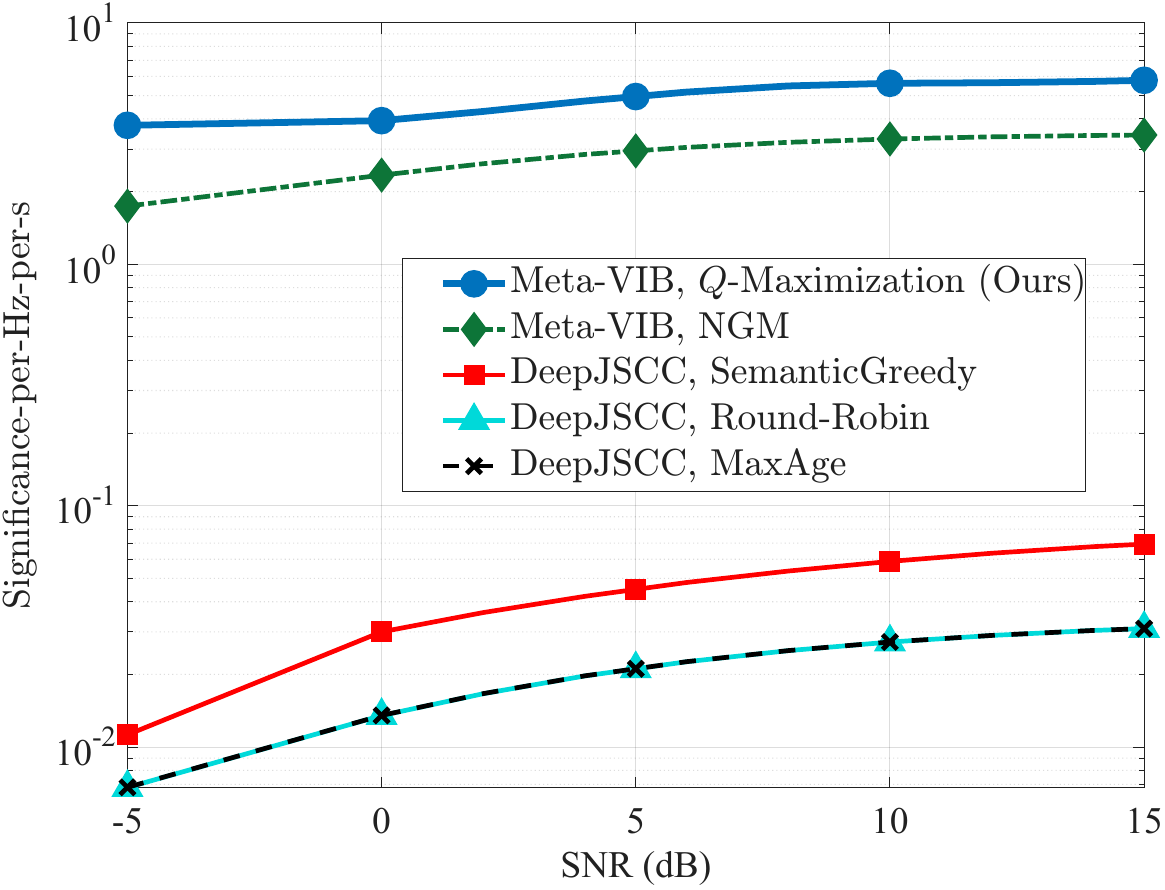}
        \label{fig:sig_per_symbol_snr}}
    \hspace{1mm}
    \subfloat[{Varying number of sensors $N$ for  channel budget $W=40$ and average SNR $\mathbb{E}[\zeta_{n,t}]=0$ dB.}]{
        \includegraphics[width=0.31\textwidth]{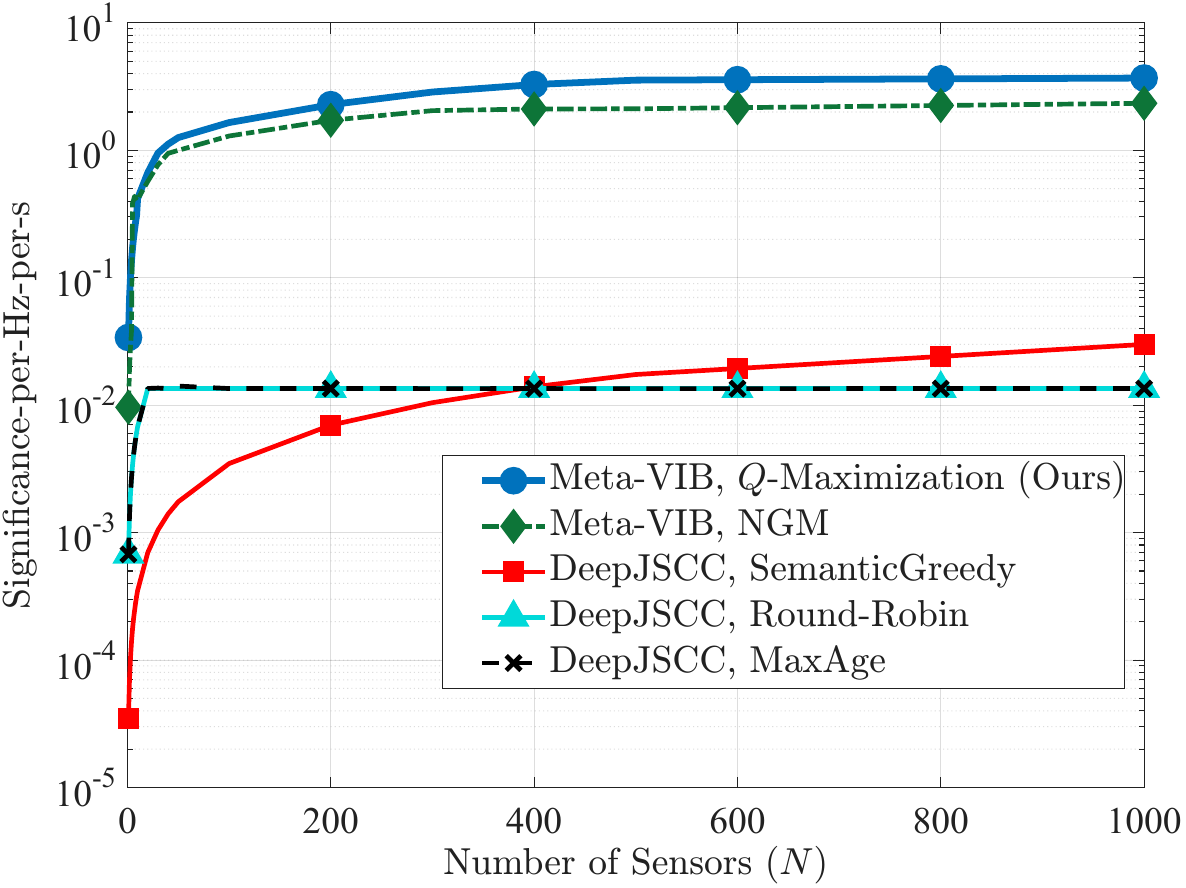}
        \label{fig:sig_per_symbol_sensors}}
    
    \caption{Significance spectrum efficiency versus the channel budget $W$, the average SNR $\mathbb{E}[\zeta_{n,t}]$, and the number of sensors $N$ for different cross-layer designs.
    }
    \label{fig:mac_evaluation}
\end{figure*}

For each camera $n$, the training loss aggregates contributions from all tracked pedestrians $i$ and all prediction horizons
$\delta=1,\ldots,\delta_{\max}$:
\begin{align}
L_n=\sum_{i}\sum_{\delta=1}^{\delta_{\max}}
\Bigl(L_{n,1}^{(i,\delta)}+0.3\,L_{n,2}^{(i,\delta)}
+0.1\,L_{n,3}^{(i,\delta)}\Bigr),
\end{align}
where $L_{n,1}^{(i,\delta)}(y,\hat{y})$ is an asymmetric classification loss
that imposes heavier penalties on safety-critical mispredictions:
\ifreport
\ignore{
\begin{align}
   \!\!\!\!\! &L_{n,1}^{(i,\delta)}(\text{dangerous},\text{safe})\!=\!100, 
     L_{n,1}^{(i,\delta)}(\text{safe},\text{dangerous})=10,\nonumber\\
   \!\!\!\!\! &L_{n,1}^{(i,\delta)}(\text{cautious},\text{safe})=20, L_{n,1}^{(i,\delta)}(\text{safe},\text{cautious})=5, 
     \nonumber\\
  \!\!\!\!\!  &
     L_{n,1}^{(i,\delta)}(\text{dangerous},\text{cautious})\!=\!15, L_{n,1}^{(i,\delta)}(\text{cautious},\text{dangerous})=5,\!\!\!\!\! \!\!\!
\end{align}
}
\begin{align}
   \!\!\!\!\! &L_{n,1}^{(i,\delta)}(\text{safe},\text{cautious})=5,
     L_{n,1}^{(i,\delta)}(\text{safe},\text{dangerous})=10,\nonumber\\
    &L_{n,1}^{(i,\delta)}(\text{cautious},\text{safe})=20,
     L_{n,1}^{(i,\delta)}(\text{cautious},\text{dangerous})=5,\nonumber\\
    &L_{n,1}^{(i,\delta)}(\text{dangerous},\text{safe})\!=\!100,
     L_{n,1}^{(i,\delta)}(\text{dangerous},\text{cautious})\!=\!15,\!\!
\end{align}
\else
\begin{align}
   \!\!\!\!\! &L_{n,1}^{(i,\delta)}(\text{safe},\text{cautious})=5,
     L_{n,1}^{(i,\delta)}(\text{safe},\text{dangerous})=10,\nonumber\\
    &L_{n,1}^{(i,\delta)}(\text{cautious},\text{safe})=20,
     L_{n,1}^{(i,\delta)}(\text{cautious},\text{dangerous})=5,\nonumber\\
    &L_{n,1}^{(i,\delta)}(\text{dangerous},\text{safe})\!=\!100,
     L_{n,1}^{(i,\delta)}(\text{dangerous},\text{cautious})\!=\!15,\!\!
\end{align}
\fi
and $L_{n,1}^{(i,\delta)}(y,\hat{y})=0$  for correct predictions. The remaining terms are
$L_{n,2}^{(i,\delta)}(y,Q_Y)=-\log Q_Y(y)$, which is a cross-entropy penalty on the reconstructed safety-label distribution, and 
$L_{n,3}^{(i,\delta)}(v,\hat{v})=\lVert v-\hat{v}\rVert_2^2$, which is an MSE penalty on the reconstructed bounding box $\hat{v}$.

We compare Meta-VIB against five physical-layer baselines: (i)
\emph{JSCC}~\cite{bourtsoulatze2019deep}, (ii) \emph{DeepJSCC}~\cite{xu2023deep}, (iii)
\emph{VIB}~\cite{alemi2016deep}, (iv) \emph{Hyper-VIB}~\cite{peng2025hyper}, and (v)
\emph{ATROC}~\cite{diao2025aligning}. 
These baselines are AoI-agnostic and non-adaptive to SNR; they are trained at fixed SNR $\zeta_{n,t}= 10$ dB and fixed AoI $\delta= 0$. Meta-VIB, by contrast, is trained over the SNR region -5 $\sim$ 20 dB and  AoI region 0 $\sim$ 500 (i.e., 0 $\sim$ 25 seconds). 
We also compare the  $Q$-Maximization algorithm against four scheduling baselines: (i) \emph{SemanticGreedy}, which prioritizes messages with higher semantic value; (ii) \emph{MaxAge}~\cite{sun2018age}, which prioritizes
sensors with the highest AoI regardless of semantic value; (iii) \emph{Round-Robin}, which cycles through sensors sequentially; and (iv) \emph{Net-Gain Maximization (NGM)} ~\cite{shisher2023learning}, which dynamically selects 
the codelength $\eta_{n,t}$, but does not satisfy the LP-priority rules of \cite{chakraborty2026advantage}. {SemanticGreedy}, {MaxAge}, and {Round-Robin} do not support codelength adaptation; their codelength is fixed at $\eta_{n,t}=2$ for all $n$ and $t$. For {NGM} and $Q$-Maximization, the codelength $\eta_{n,t}$ is selected by the algorithm.

Unless stated otherwise, the following default parameters are adopted: $m=2$, $K=8$, and $\delta_{\max}=500$. When $N$ exceeds 4, camera feeds are reused with randomized data sampling. All neural network models were trained on an NVIDIA RTX PRO 6000 Blackwell GPU. Training Meta-VIB required approximately 72 hours, while the $Q$-Maximization algorithm required approximately 3.5 hours.
\ignore{
{\blue All models were  trained on 
an NVIDIA RTX PRO 6000 Blackwell GPU. 
Training Meta-VIB required approximately $72$ hours.
The $Q$-Maximization algorithm was trained in approximately $2$ hours.}
}
\subsection{Evaluation of Per-Sample Significance}
Fig.~\ref{fig:visual_comparison} depicts the per-sample data significance
$v_{L_n,Y_{n,t+\delta}}(z)$ of the received latent $\hat{\mathbf{Z}}_{n,t}=z$
versus the instantaneous SNR $\zeta_{n,t}$ for camera $n=1$. The top and bottom rows correspond to Meta-VIB and DeepJSCC, respectively; the scalar below each panel is the per-sample significance. 
{As} SNR increases, the per-sample significance rises accordingly, and Meta-VIB consistently achieves higher semantic value than DeepJSCC. The ability to quantify the semantic value of individual messages is a key enabler for designing importance-aware semantic communication systems. Due to space limitations, 
additional visualizations of per-sample significance versus codelength $\eta_{n,t}$
and AoI $\delta$ are provided in
\ifreport
Appendix \ref{app:sec:performance_analysis}.
\else
\cite{mcdowell2026significance}.
\fi
\subsection{Evaluation of Physical-Layer Designs} \label{sec:physical_layer_analysis}

Fig.~\ref{fig:sig_per_bit_comparison} plots the average semantic significance
$\bar v_{L_n,Y_{n,t+\delta},\hat{\mathbf{Z}}_{n,t}}$ versus codelength
$\eta_{n,t}$, instantaneous SNR $\zeta_{n,t}$, and AoI $\delta$ for different physical-layer designs. 
One can observe that the average semantic value increases with codelength $\eta_{n,t}$ and SNR $\zeta_{n,t}$, while decreasing with AoI $\delta$. Moreover, Meta-VIB consistently achieves higher significance than all baselines, {as the latter neglect the AoI-dependent decay of semantic significance.} By \eqref{eq:mutual_info}, $\bar v_{L_n,Y_{n,t+\delta},\hat{\mathbf{Z}}_{n,t}}$ is upper bounded by $H_L(Y_{n,t+\delta})$, and Meta-VIB approaches this theoretical upper bound closely.
{Additional physical-layer evaluation results with SNR $\zeta_{n,t}=5$~dB and AoI $\delta=30$ are provided in
\ifreport
Appendix \ref{app:appendix_physical}.
\else
\cite{mcdowell2026significance}.
\fi
}
\vspace{-2mm}
\subsection{Evaluation of Cross-Layer Designs} \label{sec:cross_layer_analysis}

Fig.~\ref{fig:mac_evaluation} shows the semantic spectrum efficiency versus channel budget $W$, average SNR $\mathbb{E}[\zeta_{n,t}]$, and sensor count $N$ across cross-layer designs under \emph{i.i.d.} Rayleigh fading. Because we were not able to find cross-layer semantic communication designs in the prior literature, SemanticGreedy, MaxAge, and Round-Robin are each paired with DeepJSCC, while
Q-Maximization and NGM are both paired with Meta-VIB to compare the two schedulers. One can observe that the semantic spectrum efficiency decreases with the channel budget $W$ and increases with the average SNR $\mathbb{E}[\zeta_{n,t}]$. The proposed (Meta-VIB, $Q$-Maximization) consistently outperforms all baselines, achieving up to 1000 times gain in semantic spectrum efficiency at
the average SNR $\mathbb{E}[\zeta_{n,t}]=0$~dB. Moveover, the additional LP-priority-based tie-breaking step enables $Q$-Maximization to outperform NGM.
\ignore{
Fig.~\ref{fig:mac_evaluation} illustrates the semantic spectrum efficiency versus the channel budget $W$, average SNR $\mathbb{E}[\zeta_{n,t}]$, and the number of sensors $N$ for different cross-layer designs. 
Because we were not able to find cross-layer semantic communication designs in the prior literature, SemanticGreedy, MaxAge, and Round-Robin are each paired with DeepJSCC, while NGM is paired with Meta-VIB, to form the cross-layer baselines.
One can observe that the semantic spectrum efficiency decreases with the channel budget $W$, while increasing with the average SNR $\mathbb{E}[\zeta_{n,t}]$ and the number of sensors $N$. The proposed (Meta-VIB, $Q$-Maximization) design consistently outperforms all cross-layer baselines, including (Meta-VIB, NGM), confirming that learned $Q$-functions and LP-priority allocation capture multi-step dynamics beyond one-step analytical approximations.
}\textcolor{black}{Additional cross-layer evaluation results with average SNR $\mathbb{E}[\zeta_{n,t}]=5$~dB are provided in
\ifreport
Appendix \ref{app:appendix_cross}.
\else
\cite{mcdowell2026significance}.
\fi
}
\vspace{-2mm}
\section{Conclusion} \label{sec:concl}

This work proposed and validated a significance-driven cross-layer design for semantic communication. Using $L$-divergence as a per-sample measure of semantic significance, we established a rigorous information-theoretic foundation for assessing the usefulness of each data sample to the downstream inference task. Building on this foundation, we developed Meta-VIB, a physical-layer semantic encoding and inference architecture that adapts to time-varying channel conditions, symbol budgets, and data staleness without online retraining, together with the $Q$-Maximization algorithm, a lightweight significance-aware scheduler for dynamic MAC-layer channel resource allocation. Experimental results on real-world traffic-camera data show that the proposed design consistently improves semantic spectrum efficiency under varying bandwidth and SNR conditions.

\bibliographystyle{IEEEtran}
\bibliography{bibliography.bib}

\ifreport
\appendices
\section{Proof of Proposition \ref{Prop_info}} \label{App_Prop_info}
\renewcommand{\thefigure}{\thesection.\arabic{figure}}

\begin{figure*}[t]
    \centering
    \includegraphics[width=0.9\linewidth]{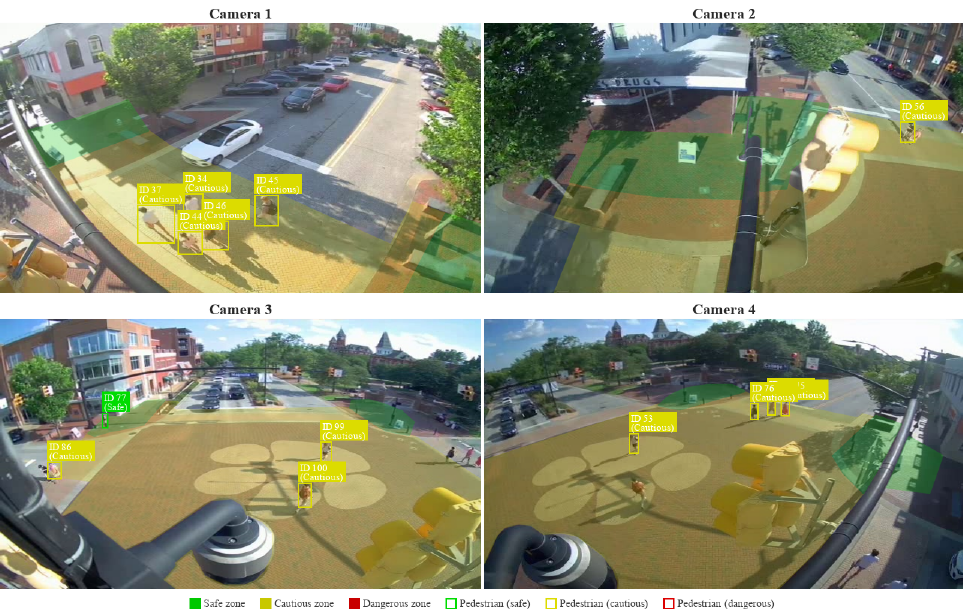}
    \caption{Annotated frames at Toomer's Corner during the red traffic light state. The active roadway is in yellow, since the traffic light is red, the pedestrians are allowed to cross and are tracked with yellow bounding boxes. The adjacent sidewalks are in green, since they are considered safe.}
    \label{app:fig:toomer}
\end{figure*}

\begin{figure*}[t]
    \centering
    \includegraphics[width=0.9\linewidth]{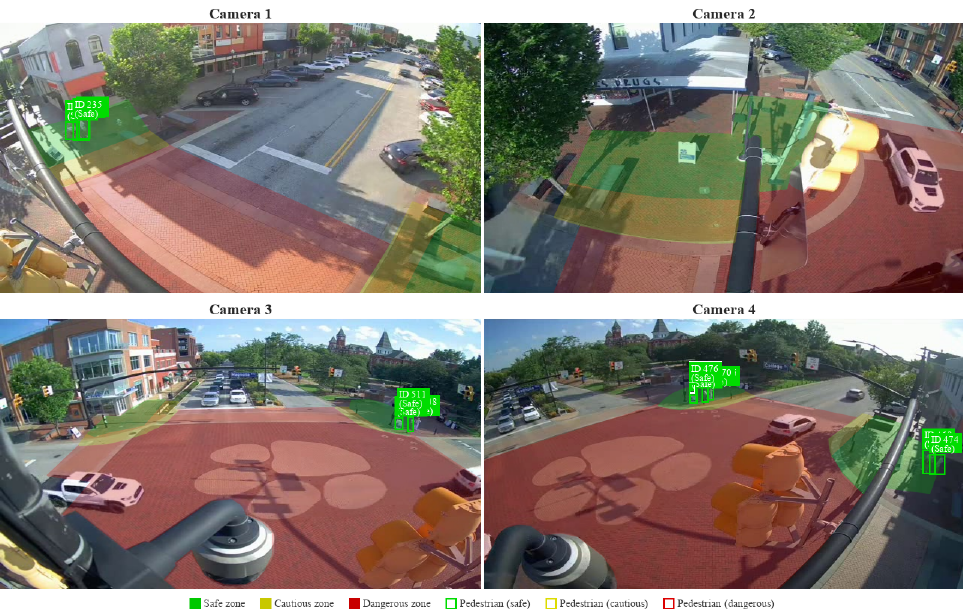}
    \caption{Annotated frames at Toomer's Corner during a single green traffic light state. The crosswalk area is in red, while the sidewalks remain in green as safe zones. Pedestrians located on the sidewalks are tracked with green bounding boxes.}
    \label{app:fig:toomer2}
\end{figure*}

\begin{figure*}[t]
    \centering
    \includegraphics[trim={0 11cm 0 10.5cm}, clip, width=0.9\linewidth]{Figures/Cameras.pdf}
    \caption{Per-sample semantic significance $v_{L_n,Y_{n,t+\delta}}(z)$ (denoted simply as $v_L$ in the figure) versus the codelength $\eta_{n,t}$, where the SNR is $\zeta_{n,t}=2$ dB and the AoI is $\delta=30$.} \label{app:fig:visual_significance_extra_codelength}
\end{figure*}

\begin{figure*}[t]
    \centering
    \includegraphics[trim={0 0 0 21cm}, clip, width=0.9\linewidth]{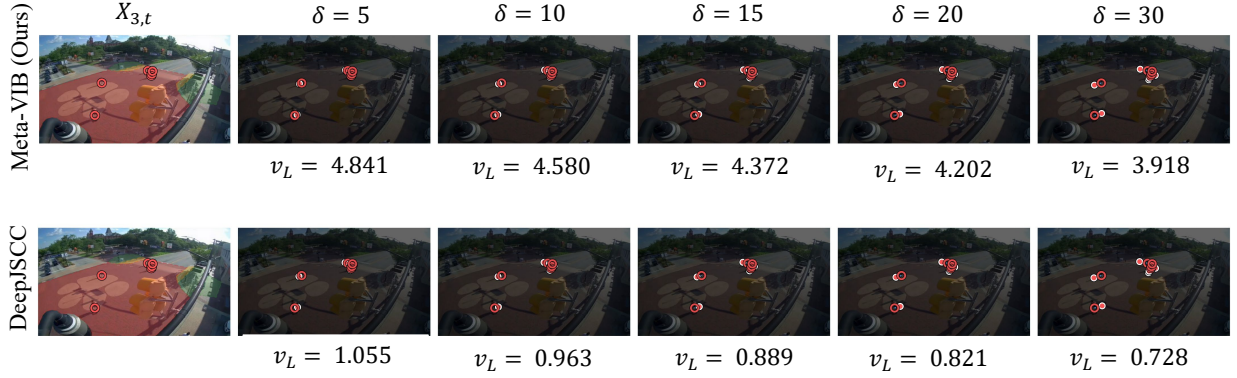}
    \caption{Per-sample semantic significance $v_{L_n,Y_{n,t+\delta}}(z)$ versus the AoI $\delta$, where the codelength is $\eta_{n,t}=2$ and the SNR is $\zeta_{n,t}=2$ dB.}
    \label{app:fig:visual_significance_extra_AoI}
\end{figure*}

\begin{figure*}[htbp]
    \centering
    \begin{subfigure}{0.33\textwidth}
        \centering
        \includegraphics[width=\linewidth]{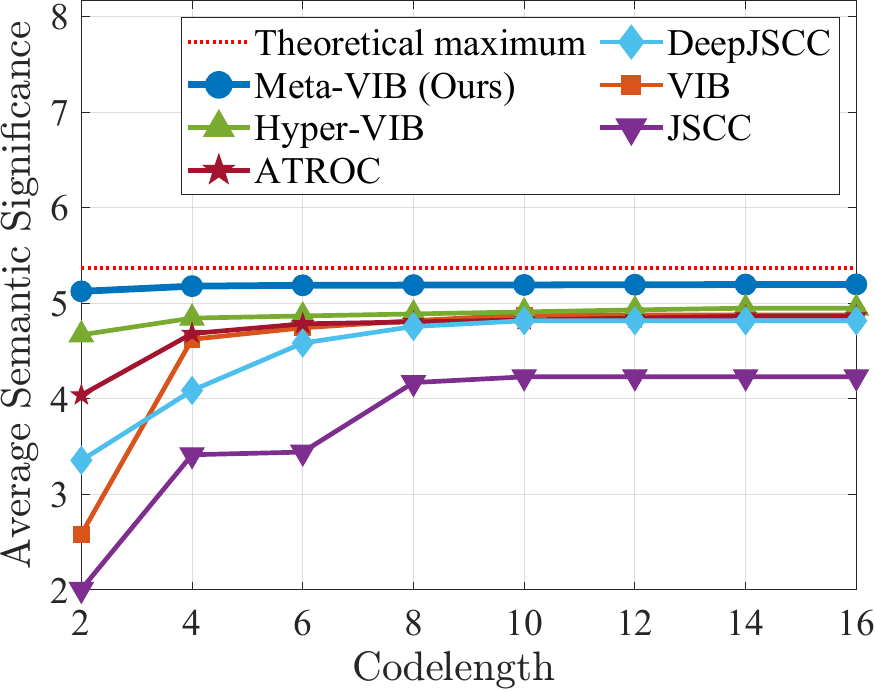}
        \caption{Varying codelength $\eta_{n,t}$ for SNR $\zeta_{n,t}= 5$ dB and AoI $\delta=30$.}
        \label{fig:sig_vs_eta}
    \end{subfigure}
    \hspace{2cm}
    \begin{subfigure}{0.325\textwidth}
        \centering
        \includegraphics[width=\linewidth]{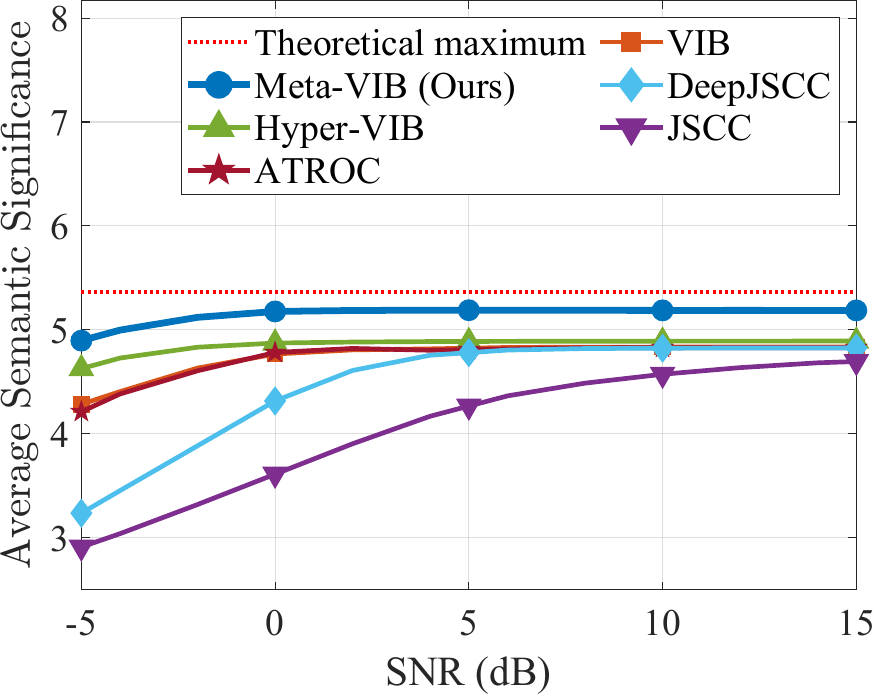}
        \caption{Varying SNR $\zeta_{n,t}$ for codelength $\eta_{n,t}=2$ and AoI $\delta=30$.}
        \label{fig:sig_vs_snr}
    \end{subfigure}
    \caption{Average semantic significance $\bar{v}_{L_n,Y_{n,t+\delta},\hat{\mathbf{Z}}_{n,t}}$ versus the codelength $\eta_{n,t}$ and the instantaneous SNR $\zeta_{n,t}$ for different physical-layer designs.} 
    \label{app:fig:physical extended}
\end{figure*}

By definition, the Bayes action associated with the prior distribution \(P_Y\) is
\begin{equation}
    a^* = a_{P_Y},
\end{equation}
and the Bayes action associated with the posterior distribution \(P_{Y|Z=z}\) is
\begin{equation}
    a^*_z = a_{P_{Y|Z=z}}.
\end{equation}
Substituting \(P=P_{Y|Z=z}\) and \(Q=P_Y\) into the definition of \(L\)-divergence yields
\begin{align}
\!\!\!D_L\!\left(P_{Y|Z=z}\,\|\,P_Y\right)
    \!&= \mathbb{E}_{Y \sim P_{Y|Z=z}}[L(Y,a_{P_Y})]
     \!\nonumber\\&-\! \mathbb{E}_{Y \sim P_{Y|Z=z}}[L(Y,a_{P_{Y|Z=z}})] \nonumber\\
    &= \mathbb{E}_{Y \sim P_{Y|Z=z}}[L(Y,a^*)]\nonumber
     \\&-\mathbb{E}_{Y \sim P_{Y|Z=z}}[L(Y,a^*_Z)].\!\!
\end{align}
The right-hand side is exactly the definition of \(v_{L,Y}(z)\) in \eqref{eq:cvi_definition}. Therefore,
\begin{equation}
    v_{L,Y}(z)=D_L\!\left(P_{Y|Z=z}\,\|\,P_Y\right).
\end{equation}
Taking the expectation with respect to \(Z\) on both sides gives
\begin{equation}
    \bar v_{L,Y,Z} = \mathbb{E}_{Z \sim P_Z}\!\left[D_L\!\left(P_{Y|Z=z}\,\|\,P_Y\right)\right].
\end{equation}
By the definition of \(L\)-mutual information, it follows that
\begin{equation}
    \bar v_{L,Y,Z} = I_L(Y;Z).
\end{equation}
By this, the proof is complete.
\section{Lagrangian Relaxation and Dual Decomposition}
\label{app:advantage}

This appendix applies Lagrangian relaxation and dual decomposition to the MAC-layer problem~\eqref{eq:mac_problem1}, yielding the per-sensor $Q$-function $Q^{*}_{n,\lambda^{*}}(\omega, \eta)$ used by the $Q$-Maximization scheduler. Let $\alpha_n(\omega) \triangleq \mathbb{P}(\omega_{n,0} = \omega)$ denote the
initial state distribution of sensor $n$, and let
$\boldsymbol{\alpha} = (\alpha_1, \ldots, \alpha_N)$ denote the collection of per-sensor initial state distributions across all $N$ sensors. The derivation proceeds in three steps and makes no assumption on the state-space cardinality or knowledge of the transition dynamics.

\ignore{
\textit{Step~1: Relaxed Problem and Dual Decomposition.}
We first relax the hard per-slot capacity constraint~\eqref{eq:global_const_mac2} to a discounted budget constraint, yielding the relaxed problem:
\begin{align}
  \max_{\pi \in \Pi} \;&
  \mathbb{E}_{\pi}\!\left[
    \sum_{t=0}^{\infty} \gamma^{t} \sum_{n=1}^{N} g_n(\omega_{n,t}, \eta_{n,t})
  \right] \label{eq:relaxed_obj} \\
  \text{s.t.} \;&
  \mathbb{E}_{\pi}\!\left[
    \sum_{t=0}^{\infty} \gamma^{t} \sum_{n=1}^{N} \eta_{n,t}
  \right] \leq \frac{W}{1-\gamma}. \label{eq:relaxed_budget}
\end{align}
}

\textit{Step~1: Relaxed Problem and Dual Decomposition.}
The hard per-slot capacity constraint~\eqref{eq:global_const_mac2}
couples all $N$ sensors, making problem~\eqref{eq:mac_problem1}
intractable. We relax it to a discounted budget constraint \eqref{eq:relaxed_budget}, yielding the relaxed problem:
\begin{subequations}\label{eq:relaxed}
\begin{align}
  \max_{\pi \in \Pi} \quad
  &\mathbb{E}_{\pi}\!\left[
    \sum_{t=0}^{\infty} \gamma^{t} \sum_{n=1}^{N}
    g_n(\omega_{n,t}, \eta_{n,t})
  \right] \label{eq:relaxed_obj} \\
  \text{s.t.} \quad
  &\mathbb{E}_{\pi}\!\left[
    \sum_{t=0}^{\infty} \gamma^{t} \sum_{n=1}^{N} \eta_{n,t}
  \right] \leq \frac{W}{1-\gamma}, \label{eq:relaxed_budget} \\
  & \eta_{n,t} \in \mathcal{A}, \quad \forall\, n,\, t,
  \label{eq:relaxed_action}
\end{align}
\end{subequations}
where 
\begin{align}
\mathcal{A} \triangleq \{0, m, 2m, \ldots, Km\}
\end{align}
is the set of available codelengths for each sensor $n$.

\ignore{
Associating the dual multiplier $\lambda \geq 0$ with constraint~\eqref{eq:relaxed_budget}, the Lagrangian is
\begin{align}
  p(\lambda)
  &= \sup_{\pi \in \Pi} \;
  \mathbb{E}_{\pi}\!\left[
    \sum_{t=0}^{\infty} \gamma^{t} \sum_{n=1}^{N}
    \Big( g_n(\omega_{n,t}, \eta_{n,t}) - \lambda\,\eta_{n,t} \Big)
  \right]
  + \frac{\lambda\,W}{1-\gamma}. \label{eq:lagrangian}
\end{align}
For fixed~$\lambda$, the Lagrangian separates across sensors. Each sensor~$n$ reduces to an independent discounted MDP with state~$\omega_{n,t}$, action~$\eta_{n,t} \in \mathcal{A}$, and $\lambda$-penalized reward $g_n(\omega_{n,t}, \eta_{n,t}) - \lambda\,\eta_{n,t}$.
}

Associating the dual multiplier $\lambda \geq 0$ with constraint~\eqref{eq:relaxed_budget} and maximizing over $\pi \in \Pi$ yields the \emph{dual function}
\begin{align}\label{eq:dual_function}
  p(\lambda)
  &\triangleq \sup_{\pi \in \Pi} \;
  \mathbb{E}_{\pi}\!\left[
    \sum_{t=0}^{\infty} \gamma^{t} \sum_{n=1}^{N}
    \Big( g_n(\omega_{n,t}, \eta_{n,t}) - \lambda\,\eta_{n,t} \Big)
  \right]\nonumber
  \\&+ \frac{\lambda\,W}{1-\gamma}.
\end{align}
Because the penalized reward decomposes additively across sensors, \eqref{eq:dual_function} decomposes into $N$ independent subproblems. Specifically, let $\pi_n=(\eta_{n,0},\eta_{n,1},\eta_{n,2},\ldots) $ denote a scheduling policy for sensor $n$. Each sensor $n$ solves an independent discounted MDP with state $\omega_{n,t}$, action $\eta_{n,t} \in \mathcal{A}$, and per-slot reward $g_n(\omega_{n,t}, \eta_{n,t}) - \lambda\,\eta_{n,t}$, given by
\begin{align}
  V^{*}_{n,\lambda}(\omega)
  &= \sup_{\pi_n}\, \mathbb{E}_{\pi_n}\!\Big[
    \sum_{t=0}^{\infty} \gamma^{t}\big(g_n(\omega_{n,t}, \eta_{n,t}) \nonumber\\&-\lambda\,\eta_{n,t}\big)
    \,\Big|\, \omega_{n,0} = \omega
  \Big],\label{eq:per_sensor_mdp}
\end{align}
where $V^{*}_{n,\lambda}(\omega)$ is the corresponding optimal value function for sensor $n$.

\ignore{
\textit{Step~2: Per-Sensor MDP Solution.}
When the transition dynamics are known, the optimal $Q$-function $Q^{*}_{n,\lambda}(\omega, \eta)$ of each per-sensor MDP satisfies the Bellman equation
\begin{align}
  Q^{*}_{n,\lambda}(\omega, \eta)
  &= g_n(\omega, \eta) - \lambda\,\eta
  + \gamma\,\mathbb{E}\big[\max_{\eta' \in \mathcal{A}} Q^{*}_{n,\lambda}(\omega', \eta') \mid \omega, \eta\big], \label{eq:bellman}
\end{align}
which is solved via standard dynamic programming methods such as value iteration or policy iteration.
}

\textit{Step~2: Solving the Per-Sensor MDP.}
For each fixed $\lambda$, the optimal $Q$-function $Q^{*}_{n,\lambda}(\omega, \eta)$ of sensor $n$ satisfies the Bellman optimality equation
\begin{align}
  Q^{*}_{n,\lambda}(\omega, \eta)
  &= g_n(\omega, \eta) - \lambda\,\eta
  + \gamma\,\mathbb{E}\!\left[\max_{\eta' \in \mathcal{A}} Q^{*}_{n,\lambda}(\omega', \eta') \,\Big|\, \omega, \eta\right]. \label{eq:bellman}
\end{align}
The value function $V^{*}_{n,\lambda}(\omega)$ and the Q-function $Q^{*}_{n,\lambda}(\omega, \eta)$ are related through
$V^{*}_{n,\lambda}(\omega) = \max_{\eta \in \mathcal{A}}
Q^{*}_{n,\lambda}(\omega, \eta)$. The Bellman optimality 
equation~\eqref{eq:bellman} holds regardless of the size of the state space or whether the transition kernel is known. In our setting, the scheduling state $\omega_{n,t}$ is high-dimensional and continuous, so $Q^{*}_{n,\lambda}$ is approximately solved using the $\lambda$-conditioned actor-critic algorithm of Section~\ref{sec:mac_ppo}. When the state space is finite and the transition kernel $P_n(\omega' \mid \omega, \eta)$ is known,~\eqref{eq:bellman} can alternatively be solved exactly by classical dynamic programming.

\ignore{
\textit{Step~3: Computing the Optimal Dual Multiplier.}
The optimal dual multiplier is obtained by solving
\begin{align}
  \lambda^{*} = \arg\min_{\lambda \geq 0}\, p(\lambda), \label{eq:dual_opt}
\end{align}
which is efficiently computed by bisection. At each candidate~$\lambda$, we solve the Bellman equation~\eqref{eq:bellman} and evaluate whether the resulting discounted resource usage meets the budget $W/(1-\gamma)$. Once $\lambda^{*}$ is found, the $Q$-values $Q^{*}_{n,\lambda^{*}}(\omega, \eta)$ from Step~2 are passed to the $Q$-Maximization scheduler.

When the state space is high-dimensional or transition dynamics are unknown, Steps~2--3 are replaced by the $\lambda$-conditioned RL architecture described in Section~\ref{sec:mac_ppo}.
}

\textit{Step~3: Optimal Dual Multiplier.}
\begin{algorithm}[t]
\caption{Bisection for the optimal dual variable $\lambda^*$}
\label{alg:bisection}
\begin{algorithmic}[0]
\Require $\lambda_l = 0$, sufficiently large $\lambda_u > 0$,
initial distribution $\boldsymbol{\alpha} = (\alpha_1, \ldots, \alpha_N)$,
tolerance $\epsilon > 0$.
\Repeat
  \State $\lambda \gets (\lambda_l + \lambda_u) / 2$.
  \State For each sensor $n$, solve~\eqref{eq:bellman} to obtain
  $Q^{*}_{n,\lambda}(\omega, \eta)$ and policy
  $\pi^*_{n,\lambda}(\omega) = \arg\max_{\eta} Q^{*}_{n,\lambda}(\omega, \eta)$.
  \State Compute $U(\lambda, \boldsymbol{\alpha}) = \sum_{n=1}^{N}
    \mathbb{E}_{\pi^*_{n,\lambda}}\!\bigl[
      \sum_{t=0}^{\infty} \gamma^t\, \eta_{n,t}
      \;\big|\; \omega_{n,0} \sim \alpha_n
    \bigr]$.
  \State \textbf{if} $U(\lambda, \boldsymbol{\alpha}) > \frac{W}{1-\gamma}$,
  $\lambda_l \gets \lambda$;
  \textbf{else}, $\lambda_u \gets \lambda$.
\Until{$\lambda_u - \lambda_l \leq \epsilon$.}
\State \textbf{return} $\lambda^* \gets \lambda$.
\end{algorithmic}
\end{algorithm}
The optimal dual multiplier $\lambda^{*}$ is obtained by solving the dual problem
\begin{align}
  \lambda^{*} = \arg\min_{\lambda \geq 0}\, p(\lambda). \label{eq:dual_opt}
\end{align}
Because $p(\lambda)$ is convex, the optimality condition is
$0 \in \partial p(\lambda^*)$, where $\partial p$ denotes the
subdifferential of the dual function $p(\lambda)$ at the point $\lambda = \lambda^*$. We solve~\eqref{eq:dual_opt} via bisection. A
subgradient of $p(\lambda)$ is given by
\begin{align}
  \frac{W}{1-\gamma} - U(\lambda, \boldsymbol{\alpha}),
  \label{eq:subgradient}
\end{align}
where 
\begin{align}
  U(\lambda, \boldsymbol{\alpha})
  \triangleq \sum_{n=1}^{N}
  \mathbb{E}_{\pi^*_{n,\lambda}}\!\left[
    \sum_{t=0}^{\infty} \gamma^t\, \eta_{n,t}
    \;\Big|\; \omega_{n,0} \sim \alpha_n
  \right]
  \label{eq:resource_usage}
\end{align}
is the total discounted resource usage under the optimal per-sensor
policies at dual variable $\lambda$. Setting~\eqref{eq:subgradient}
to zero yields the complementary slackness condition
$U(\lambda^*, \boldsymbol{\alpha}) = W/(1-\gamma)$. Since
$U(\lambda, \boldsymbol{\alpha})$ is nonincreasing in $\lambda$,
bisection efficiently finds $\lambda^*$: if
$U(\lambda, \boldsymbol{\alpha}) > W/(1-\gamma)$ the resource
usage is too high and $\lambda$ must increase; otherwise $\lambda$
must decrease. The procedure is summarized in
Algorithm~\ref{alg:bisection}. Once $\lambda^{*}$ is obtained, the
resulting $Q$-values $Q^{*}_{n,\lambda^{*}}(\omega,\eta)$ are
passed to the $Q$-Maximization scheduler.

\section{Multi-Action LP-Priority Rules for $Q$-Maximization}\label{app:lp_priority}
 
This appendix presents the multi-action LP-priority rules satisfied by the $Q$-Maximization algorithm to ensure asymptotic optimality, and describes how they are enforced through a modified dynamic programming procedure. The development follows~\cite{chakraborty2026advantage}.

\subsection{Fluid LP Relaxation}
The relaxed problem~\eqref{eq:relaxed}
can be reformulated as a Linear Program (LP) by introducing discounted
occupancy measures. Specifically, define the discounted occupancy measure as
\vspace{-1.8mm}
\begin{equation}
x_{n}(\omega, \eta) \triangleq \sum_{t=0}^{\infty} \gamma^t
\,\mathbb{P}(\omega_{n,t} = \omega,\, \eta_{n,t} = \eta).
\end{equation}
Optimizing over
$\mathbf{x} = \bigl(x_n(\omega, \eta)\bigr)_{n, \omega, \eta}$, \eqref{eq:relaxed} can be reformulated as
 the following fluid LP:
\begin{subequations}\label{eq:fluid_LP}
\begin{align}
  R_{\gamma,\mathrm{fluid}}
  &= \max_{\mathbf{x}} \;
  \sum_{n=1}^{N} \sum_{\omega}
  \sum_{\eta \in \mathcal{A}}
  g_n(\omega, \eta)\, x_n(\omega, \eta)
  \label{eq:fluid-obj} \\
  \text{s.t.} \quad
  &\sum_{n=1}^{N} \sum_{\omega}
  \sum_{\eta \in \mathcal{A}}
  \eta\, x_n(\omega, \eta)
  \leq \frac{W}{1 - \gamma},
  \label{eq:fluid-budget} \\
  &\sum_{\eta \in \mathcal{A}} x_n(\omega, \eta)
  = \alpha_n(\omega)
  \nonumber \\
  &+ \gamma \sum_{\omega', \eta'}
  P_n(\omega \mid \omega', \eta')\, x_n(\omega', \eta'),
  \ \forall\, n,\, \omega,
  \label{eq:fluid-balance} \\
  &x_n(\omega, \eta) \geq 0,
  \quad \forall\, n,\, \omega,\, \eta.
  \label{eq:fluid-nonneg}
\end{align}
\end{subequations}
where $R_{\gamma,\mathrm{fluid}}$ denotes the optimal objective value of  \eqref{eq:fluid_LP}. It is known that $R_{\gamma,\mathrm{fluid}}$ upper-bounds the optimal objective value of~\eqref{eq:mac_problem1} for every~$N$ \cite{chakraborty2026advantage}.


\subsection{Multi-Action LP-Priority Rules}

Let $\mathbf{x}^*$ denote an optimal solution
of~\eqref{eq:fluid_LP}. Given $\mathbf{x}^*$, define the set of
LP-optimal actions of sensor $n$ at state $\omega$ as
$\mathcal{A}^*_n(\omega; \mathbf{x}^*) \triangleq
\{\eta \in \mathcal{A} : x^*_n(\omega, \eta) > 0\}$.
The set $\mathcal{A}^*_n(\omega; \mathbf{x}^*)$ partitions the
sensor-state pairs $(n, \omega)$ into the following three
categories~\cite{chakraborty2026advantage}:
\begin{itemize}
\item \emph{Pure-action pairs} $\mathcal{S}^+(\mathbf{x}^*) \triangleq
\bigl\{(n, \omega) : |\mathcal{A}^*_n(\omega; \mathbf{x}^*)| = 1\bigr\}$,
with a unique LP-optimal codelength denoted $\eta^*_n(\omega)$.
\item \emph{Mixed-action pairs} $\mathcal{S}^0(\mathbf{x}^*) \triangleq
\bigl\{(n, \omega) : |\mathcal{A}^*_n(\omega; \mathbf{x}^*)| \geq 2\bigr\}$.
\item \emph{Unvisited pairs} $\mathcal{S}^-(\mathbf{x}^*) \triangleq
\bigl\{(n, \omega) : |\mathcal{A}^*_n(\omega; \mathbf{x}^*)| = 0\bigr\}$,
\end{itemize}
where $|\cdot|$ denotes set cardinality. Notice that, for every
$(n, \omega) \in \mathcal{S}^-(\mathbf{x}^*)$,
$x^*_n(\omega, \eta) = 0$ for all $\eta \in \mathcal{A}$, so the
discounted occupancy measure of these pairs satisfy
$\sum_{\eta \in \mathcal{A}} x^*_n(\omega, \eta) = 0$. Hence, pairs
in $\mathcal{S}^-(\mathbf{x}^*)$ are unreachable from the initial
state distribution $\boldsymbol{\alpha}$ in the fluid limit and
carry zero occupancy under $\mathbf{x}^*$.

Let
\begin{align}
\mathcal{X}^* \triangleq \bigl\{\mathbf{x}^* :
\mathbf{x}^* \text{ is an optimal solution
of~\eqref{eq:fluid_LP}}\bigr\}.
\end{align}
The multi-action LP-priority policy class is defined as
\[
\Pi^* \triangleq \bigcup_{\mathbf{x}^* \in \mathcal{X}^*} \Pi(\mathbf{x}^*),
\]
where $\Pi(\mathbf{x}^*)$ denotes the set of deterministic scheduling
policies that follow a priority ordering induced by $\mathbf{x}^*$
over the sensor-state-codelength triples $(n,\omega,\eta)$. At each slot $t$, a priority policy
$\pi \in \Pi(\mathbf{x}^*)$ observes the scheduling states
$\{\omega_{n,t}\}_{n=1}^N$, ranks all triples $(n, \omega_{n,t}, \eta)$ for $\eta \in \mathcal{A}$
according to a priority order induced by $\mathbf{x}^*$, and  assigns actions to the corresponding sensors in the decreasing order of priority. 
The
priority rules are specified as follows:


\begin{enumerate}
\item The triple $(n, \omega, \eta^*_n(\omega))$ with
$(n, \omega) \in \mathcal{S}^+(\mathbf{x}^*)$ receives the highest
rank, outranking all triples $(n', \omega', \eta')$ with
$(n', \omega') \in \mathcal{S}^0(\mathbf{x}^*) \cup
\mathcal{S}^-(\mathbf{x}^*)$, where $\eta^*_n(\omega)$ is the
unique LP-optimal action at the pure-action pair $(n, \omega)\in \mathcal{S}^+(\mathbf{x}^*)$.

\item At each mixed-action pair
$(n, \omega) \in \mathcal{S}^0(\mathbf{x}^*)$, the triple
$(n, \omega, \eta)$ outranks $(n, \omega, \eta')$ whenever
$\eta > \eta'$, where $\eta, \eta' \in
\mathcal{A}^*_n(\omega; \mathbf{x}^*)$ are LP-optimal actions
with $x^*_n(\omega, \eta) > 0$ and $x^*_n(\omega, \eta') > 0$. All
such triples outrank every triple $(n'', \omega'', \eta'')$ with
$(n'', \omega'') \in \mathcal{S}^-(\mathbf{x}^*)$.


\item At each pair $(n, \omega) \in \mathcal{S}^+(\mathbf{x}^*) \cup \mathcal{S}^0(\mathbf{x}^*)$, actions
$\eta \notin \mathcal{A}^*_n(\omega; \mathbf{x}^*)$ are never assigned.
\end{enumerate}
These rules specify the priority ordering across the categories
$\mathcal{S}^+(\mathbf{x}^*)$, $\mathcal{S}^0(\mathbf{x}^*)$, and
$\mathcal{S}^-(\mathbf{x}^*)$. Specifically, pure-action pairs have the highest
priority and outrank both mixed-action and unvisited pairs, and
mixed-action pairs outrank unvisited pairs. Within each
mixed-action pair, LP-optimal actions with larger codelength $\eta$
have higher priority. The rules also ensure that, for each pair
$(n,\omega)\in\mathcal{S}^+(\mathbf{x}^*)\cup
\mathcal{S}^0(\mathbf{x}^*)$, the assigned action belongs to the
LP-optimal action set $\mathcal{A}^*_n(\omega;\mathbf{x}^*)$.

The relative ordering among different pairs within
$\mathcal{S}^+(\mathbf{x}^*)$ and within
$\mathcal{S}^0(\mathbf{x}^*)$ is unspecified; any deterministic
tie-breaking rule can be used. For pairs in
$\mathcal{S}^-(\mathbf{x}^*)$, the LP-optimal solution assigns zero
occupancy to all actions and therefore does not prescribe a specific action. If such a pair appears in the finite system, any fixed action can be assigned to that pair. This arbitrary choice does not affect the fluid-limit
performance, since the pair has zero occupancy under the fluid LP
solution.


Under the mild technical conditions of~\cite{chakraborty2026advantage},
every policy in $\Pi^*$ is asymptotically optimal under the discounted
criterion. In particular, as $N$ and $W$ grow proportionally to
infinity, the total discounted significance converges to the fluid LP
upper bound $R_{\gamma,\mathrm{fluid}}$ in~\eqref{eq:fluid_LP}.

\begin{figure*}[htbp]
    \centering
    \begin{subfigure}{0.33\textwidth}
        \centering
        \includegraphics[width=\linewidth]
        {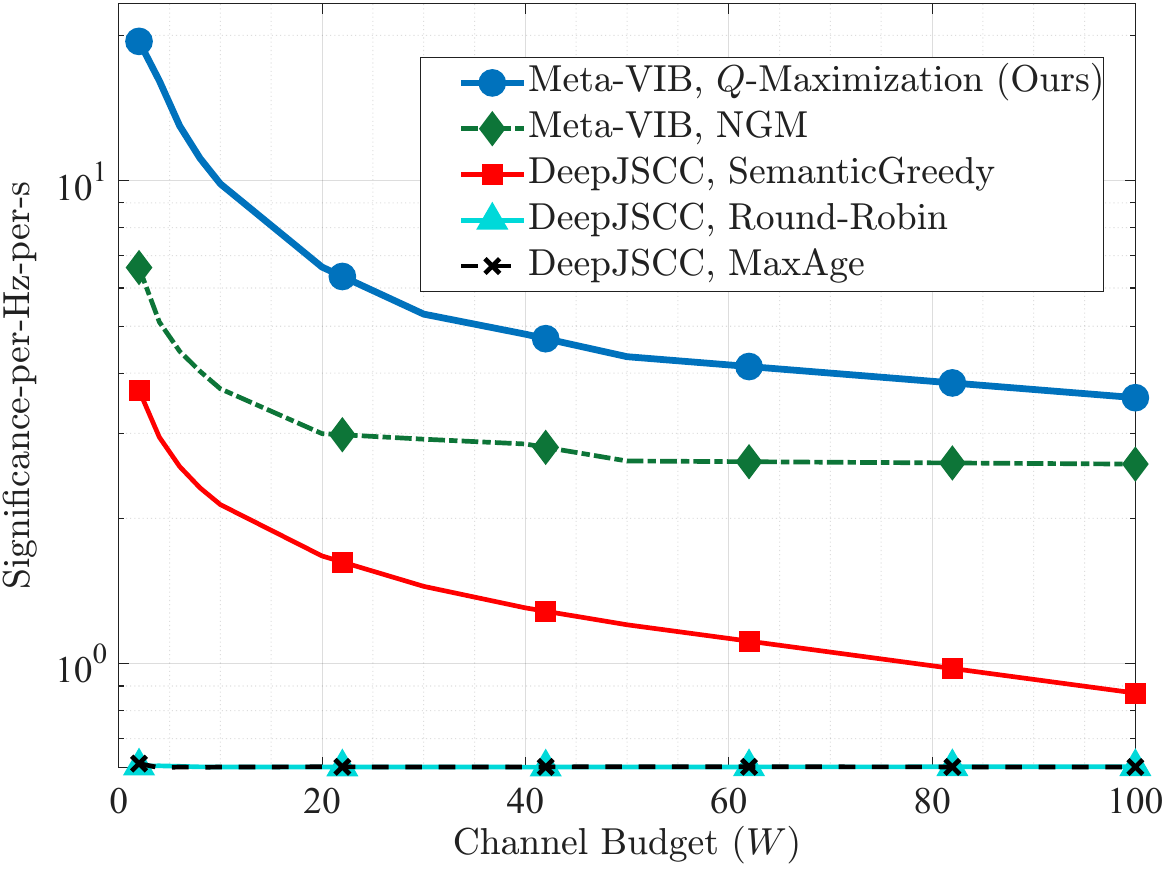}
        \caption{Varying channel budget $W$ for average SNR
        $\mathbb{E}[\zeta_{n,t}]=5$ dB and No. of sensors $N=1000$.}
        \label{fig:sig_vs_channel_budgetaoi30}
    \end{subfigure}
    \hspace{2cm}
    \begin{subfigure}{0.33\textwidth}
        \centering
        \includegraphics[width=\linewidth]
        {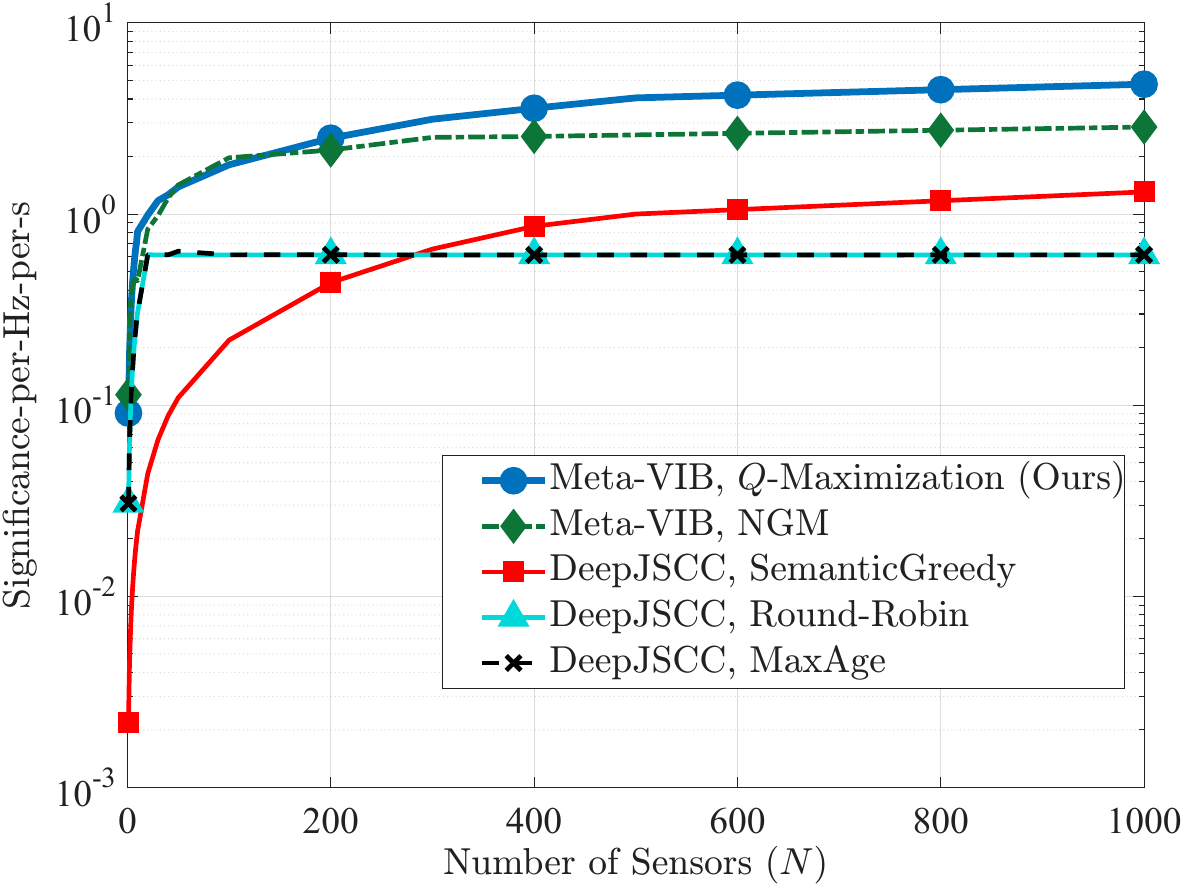}
        \caption{Varying number of sensors $N$ for channel budget $W=40$
        and average SNR $\mathbb{E}[\zeta_{n,t}]=5$ dB.}
        \label{fig:sig_vs_sensorsaoi30}
    \end{subfigure}

    \caption{Significance spectrum efficiency versus the channel budget $W$
    and the number of sensors $N$ for different cross-layer designs.}
    \label{app:fig:cross_extended}
\end{figure*}

\begin{algorithm}[t]
\caption{$Q$-Maximization via MCKP DP with LP-Priority Tie-Breaking}\label{alg:dp_tie_break}
\begin{algorithmic}[1]
\Statex \textbf{Input:} $Q$-values $\{Q_{\varphi_n}(\tilde{\omega}_{n,t}, \eta)\}_{n,\eta}$, budget $W$, codelengths $\mathcal{A} = \{0, m, 2m, \ldots, Km\}$
\Statex \textbf{Initialize:}
\State $\mathrm{dp}[0][c] \gets 0$ for all $c$;\quad $\mathrm{dp}[n][c] \gets -\infty$,\; $\mathrm{choice}[n][c] \gets \mathrm{null}$ for all $n \ge 1$
\Statex \textbf{Forward Fill:}
\For{$n = 1$ \textbf{to} $N$}
  \For{$c = 0$ \textbf{to} $W$}
    \State $\mathrm{best\_val} \gets -\infty$;\quad $\mathrm{best\_act} \gets \mathrm{null}$
    \ForAll{$\eta \in \mathcal{A}$ \textbf{with} $\eta \le c$}
      \State $\mathrm{val} \gets \mathrm{dp}[n{-}1][c - \eta] + Q_{\varphi_n}(\tilde{\omega}_{n,t},\, \eta)$
      \If{$\mathrm{val} > \mathrm{best\_val}$}
        \State $\mathrm{best\_val} \gets \mathrm{val}$;\; $\mathrm{best\_act} \gets \eta$
      \ElsIf{$\mathrm{val} = \mathrm{best\_val}$ \textbf{and} $\eta > \mathrm{best\_act}$}
        \State $\mathrm{best\_act} \gets \eta$ \hfill $\triangleright$ \textit{Tie $\to$ prefer higher cost}
      \EndIf
    \EndFor
    \State $\mathrm{dp}[n][c] \gets \mathrm{best\_val}$;\quad $\mathrm{choice}[n][c] \gets \mathrm{best\_act}$
  \EndFor
\EndFor
\Statex \textbf{Backtrack:}
\State $c \gets W$
\For{$n = N$ \textbf{downto} $1$}
  \State $\eta_{n,t} \gets \mathrm{choice}[n][c]$;\; $c \gets c - \eta_{n,t}$
\EndFor
\State \Return $\{\eta_{n,t}\}$ for all sensors $n$
\end{algorithmic}
\end{algorithm}

\subsection{Modified Dynamic Programming with LP-Priority Tie-Breaking}
\label{sec:mckp-dp}

The $Q$-Maximization algorithm belongs to the multi-action
LP-priority policy class $\Pi^*$. We establish this in two steps:
the priority rules are recovered directly from the per-sensor
$Q$-values without access to the LP solution $\mathbf{x}^*$, and
implemented through a modified dynamic programming algorithm with an additional tie-breaking step.

\subsubsection{$Q$-Value-Based Implementation of the Multi-Action LP-Priority Rules}

The $Q$-Maximization algorithm does not require the LP-optimal
solution $\mathbf{x}^*$. Instead, it implements the priority-based action selection rules
directly from the per-sensor $Q$-values. For the optimal dual
price $\lambda^*$,
\[
Q^*_{n,\lambda^*}(\omega,\eta)
\le
V^*_{n,\lambda^*}(\omega)
=
\max_{\eta'\in\mathcal{A}} Q^*_{n,\lambda^*}(\omega,\eta'),
\]
with equality exactly for the actions $\eta$ that are LP-optimal
at state $\omega$. As described in Section~\ref{sec:mac_ppo}, the learned $Q$-function
$Q_{\varphi_n}(\tilde{\omega}_{n,t},\eta)$, evaluated at the augmented state
$\tilde{\omega}_{n,t} = (\omega_{n,t}, \lambda^*)$, approximates
$Q^*_{n,\lambda^*}(\omega_{n,t}, \eta)$. Hence, at slot $t$, the $Q$-maximizing action set for
the observed pair $(n,\omega_{n,t})$ is defined as
\[
\bar{\mathcal{A}}_n(\omega_{n,t})
\triangleq
\arg\max_{\eta\in\mathcal{A}}
Q_{\varphi_n}(\tilde{\omega}_{n,t},\eta).
\]
The $Q$-Maximization algorithm enforces the LP-priority rules based
on the $Q$-maximizing action set        
$\bar{\mathcal{A}}_n(\omega_{n,t})$ as
follows~\cite{chakraborty2026advantage}:

\begin{itemize}
   \item If $\bar{\mathcal{A}}_n(\omega_{n,t})=\{\eta^*\}$, i.e.,
\(
Q_{\varphi_n}(\tilde{\omega}_{n,t},\eta^*)
>
Q_{\varphi_n}(\tilde{\omega}_{n,t},\eta),
\;\;
\forall\,\eta\in\mathcal{A}\setminus\{\eta^*\},
\)
then $(n,\omega_{n,t})$ is a pure-action pair, and action
$\eta^*$ is assigned to sensor $n$. This includes the case $\eta^*=0$, in which
sensor $n$ is left unscheduled.

    \item If $|\bar{\mathcal{A}}_n(\omega_{n,t})| \geq 2$, i.e., two or
more actions in $\mathcal{A}$ attain the maximum
$\max_{\eta} Q_{\varphi_n}(\tilde{\omega}_{n,t}, \eta)$, then
$(n, \omega_{n,t})$ is a mixed-action pair. Among the actions
in $\bar{\mathcal{A}}_n(\omega_{n,t})$, the algorithm assigns the
largest $\eta$ as long as the budget permits, and switches to the
next largest $\eta$ otherwise.


\item  At every pair $(n, \omega_{n,t})$, actions
$\eta \notin \bar{\mathcal{A}}_n(\omega_{n,t})$, i.e., actions
with $Q_{\varphi_n}(\tilde{\omega}_{n,t}, \eta) <
\max_{\eta'} Q_{\varphi_n}(\tilde{\omega}_{n,t}, \eta')$, are never
assigned as long as some action in
$\bar{\mathcal{A}}_n(\omega_{n,t})$ fits the residual budget.
\end{itemize}



When the state space is finite, under some technical conditions~\cite{chakraborty2026advantage}, the $Q$-Maximization algorithm
is asymptotically optimal for the discounted system. In
particular, as $N$ and $W$ grow proportionally to infinity, the
total discounted significance achieved by $Q$-Maximization converges
to the fluid LP upper bound $R_{\gamma,\mathrm{fluid}}$.

For high-dimensional continuous state spaces, the per-sensor $Q$-functions $Q_{\varphi_n}(\tilde{\omega}_{n,t},\eta)$ are learned directly from data using the
$\lambda$-conditioned actor-critic training procedure described in
Section~\ref{sec:mac_ppo}. The $Q$-Maximization algorithm then uses these learned $Q$-values to
implement the priority rules through the modified dynamic
programming procedure described next.

\subsubsection{Dynamic Programming Algorithm for $Q$-Maximization}
Standard dynamic programming~\cite{kellerer2004multidimensional}
solves the Multiple-Choice Knapsack Problem (MCKP)~\eqref{eq:mckp}
in two steps. The first step is the forward fill, which computes, for
each sensor and each budget level, the maximum total
$Q$-value that can be achieved. The second step is backtracking, which
uses the recorded decisions to recover the codelength assigned to each
sensor. Algorithm~\ref{alg:dp_tie_break} follows this same two-step
structure, with a single modification to the tie-breaking rule in the
forward fill step.
When multiple codelengths yield the same optimal $Q$-value for a
mixed-action pair, standard MCKP dynamic programming breaks ties
arbitrarily, which can leave budget unused and lead to a suboptimal
long-run state distribution as $N$ and $W$ grow proportionally to
infinity~\cite{chakraborty2026advantage}. To enforce the multi-action LP-priority rules, which give priority
to the larger codelengths among the maximizers at a mixed-action
pair, Algorithm~\ref{alg:dp_tie_break} modifies only the forward
fill by keeping the largest tied codelength; the
backtrack step remains unchanged. The two steps are detailed
below.

\textit{Step~1: Forward fill.}
The forward fill constructs the dynamic programming table
$\mathtt{dp}[n][c]$ for $n=1,\ldots,N$ and $c=0,\ldots,W$. The entry
$\mathtt{dp}[n][c]$ is the maximum total $Q$-value achievable by
assigning one codelength to each of the first $n$ sensors using at most
$c$ channel symbols. For sensor $n$ and budget level $c$, the algorithm evaluates each
codelength $\eta \in \mathcal{A}$ with $\eta \le c$ and computes
\(
\mathtt{val} = \mathtt{dp}[n{-}1][c-\eta] +
Q_{\varphi_n}(\tilde{\omega}_{n,t}, \eta).
\)
If $\mathtt{val}$ strictly exceeds the current best value, the table
entry and the selected codelength are updated. If $\mathtt{val}$
equals the current best value, the larger of the two tied
codelengths is kept. This tie-breaking is the only
difference from standard MCKP dynamic
programming~\cite{kellerer2004multidimensional}. The selected
codelength is recorded in $\mathtt{choice}[n][c]$.

\textit{Step~2: Backtrack.}
The backtrack recovers the codelength allocation
$\boldsymbol{\eta}_t$ by tracing through the $\mathtt{choice}$
table. Starting from $c = W$, the algorithm iterates from sensor
$N$ down to sensor $1$. At each sensor $n$, it reads
$\eta_{n,t} = \mathtt{choice}[n][c]$ and updates the residual
budget to $c - \eta_{n,t}$. The complexity of Algorithm~\ref{alg:dp_tie_break} is $O(NWK)$,
in the same order with standard MCKP dynamic programming. 

\subsection{Integration with the RL Architecture}
In our setting, the scheduling
state $\omega_{n,t}$ is high-dimensional and continuous. The
$\lambda$-conditioned RL architecture of
Section~\ref{sec:lambda_rl} learns
$Q_{\varphi_n}((\omega_{n,t}, \lambda), \eta)$ directly from data. At
deployment, the learned $Q$-values are passed to
Algorithm~\ref{alg:dp_tie_break}, which recovers the partition from
the $Q$-value ties and applies the priority-based tie-breaking as
described above.
\section{Pedestrian Risk Zones and Visualizations}
\label{sec:appendix_visuals}
\setcounter{figure}{0}
Toomer's Corner operates under an all-stop pedestrian crossing scheme: during the \textit{all-stop phase}, every vehicle signal is red and pedestrians are permitted to cross in any direction, including diagonally; during the \textit{vehicle phase}, vehicles actively traverse the intersection and pedestrians remain on the sidewalks. We segment the spatial field of view into three semantic risk zones whose labels depend jointly on pedestrian position and the instantaneous traffic phase. The \textit{safe} zone encompasses off-road infrastructure, including sidewalks and pedestrian waiting areas, and persists regardless of phase. The \textit{cautious} zone covers the painted crosswalk and curb-adjacent transition areas during the all-stop phase, when pedestrians are permitted to cross but remain physically vulnerable to non-compliant vehicles. The \textit{dangerous} zone covers the active vehicle roadway and absorbs the crosswalk surfaces during the vehicle phase.

The zone definitions in Section~\ref{sec6.1} are a training-time redefinition of conventional pedestrian-safety semantics. Under standard usage, \textit{dangerous} would denote a verified crash or near-miss event; we instead label pre-crash configurations as the dangerous-event ground truth, identifying as dangerous any pedestrian in the crosswalk during the vehicle phase, with \textit{cautious} and \textit{safe} labels assigned analogously based on the instantaneous traffic phase and pedestrian position. This redefinition is necessary because real crashes and near-misses are statistically rare, producing a label distribution far too sparse to train a semantic communication model on directly. Following an increasingly common practice in transportation safety research, we instead generate \textit{unsafe} supervision signals from normal traffic interactions, yielding a dense and learnable label distribution. We emphasize that this redefinition applies only at training; at deployment, the trained model serves the conventional safety objective of detecting real-world rare dangerous events as they occur. The asymmetric loss in~\eqref{eq:loss} reflects this safety prioritization, imposing the heaviest penalty on a \textit{dangerous}-to-\textit{safe} misprediction so that the receiver strictly avoids misclassifying actively vulnerable pedestrians as secure.

\subsection{Visualizations Across Traffic Phases}
Figures~\ref{app:fig:toomer} and~\ref{app:fig:toomer2} illustrate annotated frames captured at Toomer's Corner under the two traffic phases. Figure~\ref{app:fig:toomer} depicts the all-stop phase, which has all traffic lights red: the crosswalk surfaces are overlaid in yellow as the \textit{cautious} zone, and pedestrians traversing this region are tracked with yellow bounding boxes labeled \textit{Cautious}. Figure~\ref{app:fig:toomer2} depicts the vehicle phase, which has at least one traffic signal green: the crosswalk and adjacent vehicle roadway are overlaid in red as the \textit{dangerous} zone, while pedestrians waiting on the sidewalks are tracked with green bounding boxes labeled \textit{Safe} and reside within the \textit{safe} zone. Together, these visualizations show that the semantic risk of a given camera frame is highly dynamic and conditioned on the instantaneous traffic phase, motivating the per-sample significance metric that governs the proposed scheduling framework.

\section{Additional Evaluation of Per-Sample Significance} \label{app:sec:performance_analysis}
\textcolor{black}{Figs.~\ref{app:fig:visual_significance_extra_codelength} and \ref{app:fig:visual_significance_extra_AoI} provide a broader analysis of the per-sample significance $v_{L_n,Y_{n,t+\delta}}(z)$ with respect to the codelength $\eta_{n,t}$ and the AoI $\delta$, respectively. \textcolor{black}{In these visualizations, empty colored circles represent the true positions and safety labels of pedestrians, while solid colored circles represent the estimated positions and safety labels reconstructed by the receiver's decoder. The color of each circle indicates the assigned safety risk level: red for dangerous, yellow for cautious, and green for safe.} These results confirm that the performance gains of Meta-VIB are not isolated to specific SNR regimes; instead, Meta-VIB consistently achieves higher semantic values than DeepJSCC, which suffers from substantially higher misclassification of trajectories error across all operating conditions. Notably, the advantage is most pronounced in the low-$\eta$ regime and at high $\delta$, where the resource scarcity and information staleness make the significance-driven compression of Meta-VIB indispensable for maintaining communication utility.}

\subsection{Physical-Layer Significance Evaluation}
\label{app:phy_significance_eval}

\paragraph{Offline preparation.} Before deployment, three quantities are
precomputed and stored at the receiver:
(i)~the prior-optimal action
$\hat a^* = \arg\min_{a}\,
\mathbb{E}_{Y\sim\widehat P_{Y_{n,t+\delta}}}\![L_n(Y,a)]$,
obtained from the empirical training-set label distribution
$\widehat P_{Y_{n,t+\delta}}$;
(ii)~the Meta-VIB parameters $(\phi_n^*,\theta_n^*,\psi_n^*)$ from the
three-phase procedure of Section~\ref{sec:meta_vib}; and
(iii)~the $Q$-function $Q_{\varphi_n}(\tilde\omega,\eta)$ of
Section~\ref{sec:lambda_rl}, which was trained using the expected significance $g_n(\omega,\eta)$ and dual multiplier $\lambda$.

\paragraph{Online per-sample data significance evaluation.} At slot $t$, given the
realization $\hat{\mathbf Z}_{n,t}=z$, the receiver runs a single forward
pass of the trained decoder to produce the empirical posterior
$\widehat P_{Y_{n,t+\delta}\mid S_{n,t}=z}$. The posterior-optimal action
follows from~\eqref{eq_posterior_action} as
\begin{align}\label{eq:posterior_action_emp}
\hat a^*_z = \arg\min_{a}\,
\mathbb{E}_{Y\sim\widehat P_{Y_{n,t+\delta}\mid S_{n,t}=z}}\![L_n(Y,a)],
\end{align}
and the per-sample significance is then evaluated directly via~\eqref{eq:cvi_definition}:
\begin{align}\label{eq:vL_per_sample}
v_{L_n,Y_{n,t+\delta}}(z)
=\mathbb{E}_{Y\sim\widehat P_{Y_{n,t+\delta}\mid S_{n,t}=z}}\!\!
\bigl[L_n(Y,\hat a^*) - L_n(Y,\hat a^*_z)\bigr].
\end{align}
produces the per-sample values plotted in
Figs.~\ref{fig:visual_comparison},
\ref{app:fig:visual_significance_extra_codelength}, and
\ref{app:fig:visual_significance_extra_AoI}.

\paragraph{MAC-layer utilization.} Once the physical layer is frozen, the
scheduler operates entirely from receiver-side information. At each slot
$t$, the receiver evaluates the trained
$Q_{\varphi_n}(\tilde\omega_{n,t},\eta)$ once for every $\eta\in\mathcal A$
and feeds the Q-values 
to Algorithm~\ref{alg:dp_tie_break}, which
outputs the codelength allocation $\boldsymbol\eta_t$ in $O(NWK)$ time. 

\section{Additional Evaluation of Physical-Layer Designs} 
\label{app:appendix_physical}

\textcolor{black}{Fig.~\ref{app:fig:physical extended} plots the average semantic significance $\bar v_{L_n,Y_{n,t+\delta},\hat{\mathbf{Z}}_{n,t}}$ versus codelength $\eta_{n,t}$ and instantaneous SNR $\zeta_{n,t}$ for a more favorable operating scenario. One can observe that the performance trends remain qualitatively consistent with the previous analysis in Section~\ref{sec:physical_layer_analysis}. The average semantic value increases with codelength $\eta_{n,t}$ and SNR $\zeta_{n,t}$. Although the performance gain over the baselines is less pronounced in this regime, Meta-VIB consistently achieves higher significance than all considered baselines and approaches the upper bound $H_L(Y_{n,t+\delta})$ closely. This gap reflects a key limitation of the baselines, which are trained for a fixed SNR of $10$ dB and a fixed AoI of $0$, and therefore do not explicitly account for SNR variations or age-induced degradation. In contrast, Meta-VIB is trained over a broad operating region with SNR from $-5$ dB to $20$ dB and AoI from $0$ to $500$ slots, corresponding to $0$ to $25$ seconds.}


\section{Additional Evaluation of Cross-Layer Designs} 
\label{app:appendix_cross}

\textcolor{black}{In addition to the cross-layer evaluations presented in Section~\ref{sec:cross_layer_analysis}, Fig.~\ref{app:fig:cross_extended} illustrates the semantic spectrum efficiency versus the channel budget $W$ and the number of sensors $N$ considering a more favorable operating scenario: average SNR $\mathbb{E}[\zeta_{n,t}]=5$ dB. One can observe that the semantic spectrum efficiency decreases with the channel budget $W$ while increasing with the number of sensors $N$. Our design (Meta-VIB, $Q$-Maximization) consistently outperforms all cross-layer baselines because it dynamically adapts the codelength allocation to the semantic value and channel state, whereas fixed-codelength baselines cannot increase the codeword length when
the instantaneous channel capacity is insufficient for reliable inference.
Although NGM also supports dynamic codelengths, it lacks the asymptotic
optimality property of $Q$-Maximization and therefore achieves lower semantic
spectrum efficiency.}

\else
\fi

\end{document}